\documentclass{emulateapj}

\shorttitle{ULXs in the Interacting Systems}
\shortauthors{Yoshida et al.}

\begin{document}

\title{Long-Term Spectral Variations of Ultraluminous X-ray Sources \\in the interacting galaxy systems M\,51 and NGC\,4490/85}

\author{
Tessei~Yoshida\altaffilmark{1,2},
Ken~Ebisawa\altaffilmark{1},
Kyoko~Matsushita\altaffilmark{2},
Masahiro~Tsujimoto\altaffilmark{1},
and Toshihiro~Kawaguchi\altaffilmark{3}
}

\email{yoshida.tessei@ac.jaxa.jp}

\altaffiltext{1}{Japan Aerospace Exploration Agency, Institute of Space and Astronautical Science, 3-1-1 Yoshinodai, Chuo, Sagamihara, Kanagawa 252-5210, Japan}
\altaffiltext{2}{Department of Physics, Tokyo University of Science, 1-3 Kagurazaka, Shinjuku, Tokyo 162-8601, Japan}
\altaffiltext{3}{Center for Computational Sciences, University of Tsukuba, 1-1-1 Tennodai, Tsukuba, Ibaraki 305-8577, Japan}

\begin{abstract}

Variable ultraluminous X-ray sources (ULXs), which are considered to be black hole binaries (BHBs),
are known to show state transitions similarly to Galactic BHBs.
However, the relation between the ULX states and the Galactic BHB states is still unclear 
primarily due to less well-understood behaviors of ULXs in contrast to the Galactic BHBs.
Here, we report a statistical X-ray spectral study of 34 energy spectra from seven bright ULXs in the interacting galaxy systems M\,51 and NGC\,4490/85,
using archive data from multiple {\it{Chandra}} and {\it{XMM-Newton}} observations spanning for a few years.
In order to compare with Galactic BHB states, we applied representative spectral models of BHBs;
a power-law (PL), a multi-color disk black body (MCD), and a slim disk model to all the ULX spectra.
We found a hint of a bimodal structure in the luminosity distribution of the samples,
suggesting that ULXs have two states with typical luminosities of 3--6$\times$10$^{39}$ and 1.5--3$\times$10$^{39}$~ergs~s$^{-1}$.
Most spectra in the brighter state are explained by the MCD or the slim disk model, whereas those in the fainter state are explained by the PL model.
In particular, the slim disk model successfully explains the observed spectral variations of NGC\,4490/85 ULX-6 and ULX-8 by changes of the mass accretion rate
to a black hole of an estimated mass of $<$40~$M_{\odot}$.
From the best-fit model parameters of each state, we speculate that the brighter state in these two ULXs corresponds to the brightest state of Galactic BHBs,
which is often called the ``apparently standard state''.
The fainter state of the ULXs has a PL shaped spectrum, but the photon index range is much wider than that seen in any single state of Galactic BHBs.
We thus speculate that it is a state unique to ULXs.
Some sources show much fainter and steeper spectra than the faint state, which we identified as another state.
\end{abstract}

\keywords{accretion, accretion disks --- black hole physics --- galaxies: individual (M\,51, NGC\,4490, NGC\,4485) --- X-rays: binaries}

\section{Introduction}

Ultraluminous X-ray sources (ULXs) are off-nuclear point-like sources detected in the X-ray bandpass with luminosities of $\ge$$10^{39-41}$~ergs~s$^{-1}$;
see \citet{liu05}, \citet{fab06} for a review.
Most variable ULXs are considered to be black hole binaries (BHBs) based on observational characteristics similar to Galactic BHBs,
such as short- and long-term variations \citep[e.g.][]{ptak99b}, thermal disk emission \citep[e.g.][]{max00,fen10}, and state transitions \citep[e.g.][]{kubo01}.
The large luminosity of ULXs suggests the black hole (BH) mass ($M$) to be much larger than that of Galactic BHs ($\sim$10~$M_{\odot}$) under the assumption
that the radiation is spherically symmetric and the observed luminosity does not exceed the Eddington luminosity
$L_{\rm{Edd}}$~$=$~1.5$\times 10^{38}$($M$/$M_{\odot}$)~ergs~s$^{-1}$.

Three competing interpretations have been proposed for ULXs.
The first is that ULXs contain a so-called intermediate mass BH with a mass of 100--1000~$M_{\odot}$ \citep{mil04}.
In fact, a large mass of $\gtrsim$700~$M_{\odot}$ is suggested for M\,82 X-1 based on its large luminosity \citep{mat01,kaa01}.
The second is that ULXs are BHs with a mass comparable to or slightly larger than
that of Galactic BHs of $<$40~$M_{\odot}$ \citep[so-called stellar mass BHs;][]{ebi03,oka06}.
The super-Eddington luminosity is interpreted as a consequence of the ULXs having a slim disk \citep{abr88} rather than the standard disk \citep{sha73}.
In the state expressed by the slim disk model,
which is a stable solution for very high mass accretion rates of $\gtrsim$8$\times 10^{18}$($M$/$M_{\odot}$)~g~s$^{-1}$,
the emission is moderately collimated toward the direction normal to the disk so that the luminosity can be super-Eddington \citep[e.g.][]{ohs05}.
The third is the beaming model with a high collimation \citep{king01},
although the scenario may have difficulties in explaining extended photoionized nebulae found around some ULXs.
For example, \citet{kaa04} studied the Holmberg\,II ULX within an extended nebula.
The nebula emits He~II 4686~$\rm{\AA}$ line relatively isotropically, which is produced by the X-ray photoionization from the central ULX.
Thus, the extended nebula is unlikely excited by highly collimated radiation.

Galactic BHBs are known to show transitions among several states \citep{esin97}.
The most three well-established states are the low-hard state (LHS), the high-soft state (HSS), and the very high state (VHS).
In general, the LHS is the faintest state \citep[0.01--0.04~$L_{\rm{Edd}}$,][]{mac03},
which shows a power-law (PL) spectrum with a photon index ${\it{\Gamma}}$~=~1.5--1.9 \citep{esin97}.
The HSS is generally brighter than the LHS, and is a thermal state with a convex-shaped spectrum.
The spectrum can be fitted by a multi-color disk black body (MCD) model \citep{pri81},
which is an approximation for the standard disk spectra.
The VHS usually shows an X-ray luminosity comparable or higher than that of the HSS and a steep PL-like spectrum \citep[typically ${\it{\Gamma}}$~$\sim$~2.5,][]{esin97}.
The spectrum is considered to be a mixture of a weak disk component and a strong Comptonized component.

In addition to these well-established three states, yet another state --- the apparently standard state (ASS) --- was proposed by \citet{kubo04}.
The ASS is a thermal state, which is brighter ($L_{\rm{X}}$~$\gtrsim$~0.3~$L_{\rm{Edd}}$) than the other states.
At least two Galactic BHBs are reported to show this state: XTE\,J1550--564 \citep{kubo04} and 4U\,1630--47 \citep{abe05}.
The spectral shape in the ASS is better represented by a slim disk model than the MCD model.
If the spectra in this state are fitted with the MCD model, the following three anomalies appear \citep{kubo04,abe05}:
(i) Excess emission is found below 5~keV and above 10~keV.
(ii) The innermost disk temperature is higher than that in the HSS.
(iii) The innermost disk radius tends to decrease with the increasing the disk temperature.
In the HSS, in contrast, the innermost disk temperature is rather constant \citep[e.g.][]{ebi94}.
Some of these anomalies can be solved either by using the slim-disk model instead of the MCD model \citep{kubo04,abe05},
or by introducing spectral hardening to the MCD model as a function of accretion rate \citep{dav06,mcc07}.
Although it is yet unclear if the ASS should be considered to be an independent state or an extension of the HSS,
we treat the four states (LHS, HSS, VHS, and ASS) as different states in this paper.

Similarly to the Galactic BHBs, some ULXs are known to show state transitions in timescales of months to years between two different spectral states.
In the case of IC\,342 sources\,1 and 2, \citet{kubo01} dubbed them the ``hard'' and the ``soft'' state, which respectively has a spectrum with a PL and a convex shape.
The latter is brighter than the former by from a factor of a few to one order of magnitude.
However, the relation between the spectral states of ULXs and those of Galactic BHBs is still unclear.
Observational clues will be obtained by comparing the state transitions,
including their frequencies and time scales of transitions.

ULXs are less well frequently monitored than Galactic BHBs, and their time variation is less understood.
Obviously, increasing the samples of ULXs is a key to conduct statistical studies of state transitions.
Nearby ($<$5~Mpc) galaxies and interacting galaxy systems at a moderate distance (5--10~Mpc) are two suitable laboratories for this purpose.
ULXs in nearby galaxies are bright in flux and can be examined in detail.
About 15 ULXs were studied within 5~Mpc,
which includes NGC\,1313 X-1 and 2 \citep{miz07}, IC\,342 sources\,1 and 2 \citep{kubo01},
Holmberg\,II X-1 \citep{goa06}, M\,81 X-9 \citep{tsuno06}, M\,82 X-1 \citep{miy09}, and NGC\,5204 X-1 \citep{rob06}.

We focus on the other laboratory; ULXs in interacting galaxy systems.
They are more distant and fainter in flux than those in nearby galaxies.
However, the average number of ULX per galaxy is much larger, enabling us to monitor a large number of samples at the same distance simultaneously.
The number of ULXs in nine interacting systems at 20--100~Mpc \citep{bra07} averages to be $>$6~ULXs~system$^{-1}$.

Within 10~Mpc, the M\,51 system and the NGC\,4490 and NGC\,4485 system (hereafter called NGC\,4490/85) host the largest number of ULXs.
The former at $\sim$8.4~Mpc \citep{fel97} hosts nine, while the latter at $\sim$8~Mpc \citep{deVau76} hosts eight.
Both systems have been observed several times in X-rays and thus are suitable to study long-term variations of multiple ULXs at a time.
The X-ray data from these two systems have not been uniformly analyzed yet.
In this paper, we apply physical models to the spectra of these ULXs and
give interpretations to the observed long-term spectral changes, in comparison with the Galactic BHB state transitions.

\begin{deluxetable}{ccccc}
\tablecaption{Observation Log.\label{table1}}
\tablewidth{0pt}
\tablecolumns{5}
\tablehead{
Data & Observatory & ObsID & Date & $t_{\rm{exp}}$\tablenotemark{a} \\
label & & & & (ks)
}
\startdata
\multicolumn{2}{l}{M\,51} & & & \\
\cline{1-5}
C1 & {\it{Chandra}} & 354 & 2000-06-20 & 14.9 \\
C2 &  & 1622 & 2001-06-23 & 26.8 \\
C3 &  & 3932 & 2003-08-07 & 48.0 \\
X1 & {\it{XMM-Newton}} & 0112840201 & 2003-01-15 & 20.7/19.0 \\
X2 &  & 0212480801 & 2005-07-01 & 49.0/47.3 \\
X3 &  & 0303420101 & 2006-05-20 & 52.5/52.2 \\
X4 &  & 0303420201 & 2006-05-24 & 36.6/34.9 \\
\tableline
\multicolumn{2}{l}{NGC\,4490/85} & & & \\
\cline{1-5}
C1$^{\prime}$ & {\it{Chandra}} & 1579 & 2000-11-03 & 19.5 \\
C2$^{\prime}$ &  & 4725 & 2004-07-29 & 38.5 \\
C3$^{\prime}$ &  & 4726 & 2004-11-20 & 39.6 \\
X1$^{\prime}$ & {\it{XMM-Newton}} & 0112280201 & 2002-05-27 & 17.4/11.9
\enddata
\tablenotetext{a}{ACIS exposure for the {\it{Chandra}} observations and EPIC MOS (left) and EPIC pn (right) exposure for the {\it{XMM-Newton}} observations.}
\end{deluxetable}

\begin{figure*}
\plotone{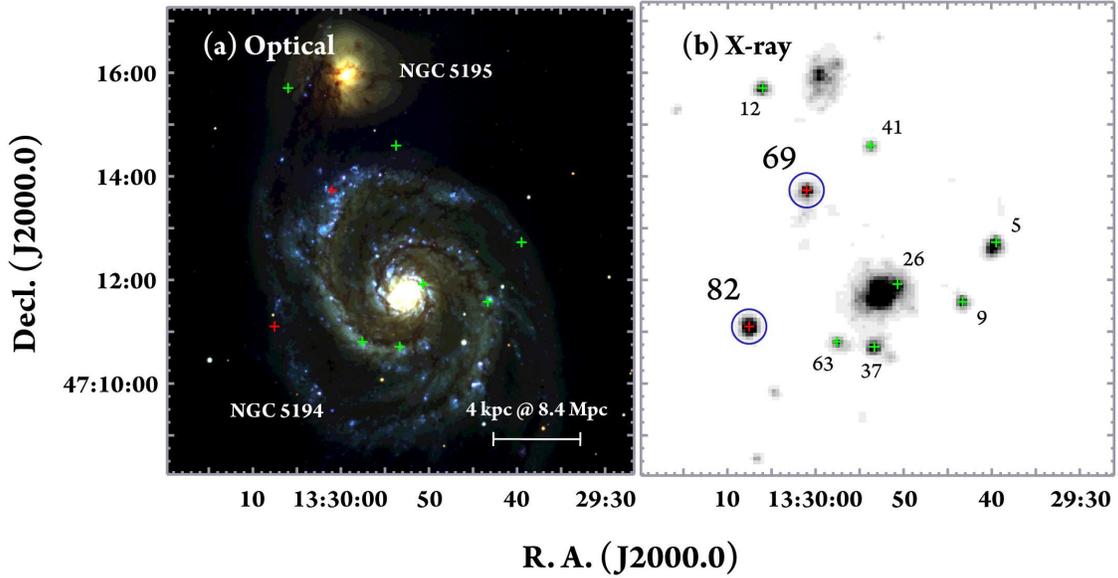}
\caption{Optical and X-ray images of M\,51.
(a) SDSS three color image.
Red, green, and blue respectively represents the {\it{r}}-, {\it{g}}-, and {\it{u}}-band intensity.
Pluses show positions of the nine ULXs in \citet{dew05};
the two brightest sources, focused in this study, are shown in red and the remainders are in green.
(b) Smoothed {\it{XMM-Newton}} MOS-1 image in the 0.5--8.0~keV band.
Solid blue circles indicate the source extraction region.
\label{fig1}}
\end{figure*}

\begin{figure*}
\plotone{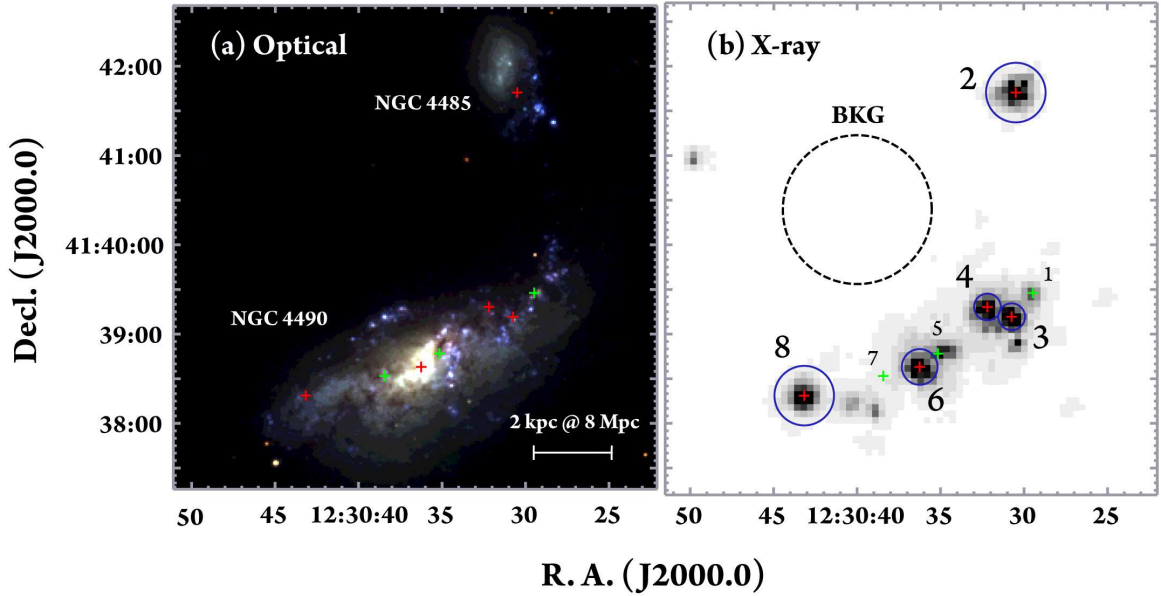}
\caption{Same with Figure~\ref{fig1}, but for NGC\,4490/85.
Pluses show positions of the eight ULXs in \citet{fri08}.
The dashed circle in (b) indicates the common background region.
\label{fig2}}
\end{figure*}

\section{Objects}

\subsection{M\,51}
M\,51 is one of the closest interacting galaxy systems, which is composed of two spiral galaxies, NGC\,5194 and NGC\,5195.
The sizes of NGC\,5194 and NGC\,5195 are 27 and 14~kpc from the isophotal diameters 11\farcm2 and 5\farcm8 \citep{deVau91}, respectively.
The total mass of the M\,51 system is $\sim$1.5$\times 10^{11}$~$M_{\odot}$ \citep{kuno97},
and the mass of NGC\,5194 is larger than that of NGC\,5194 by a factor of a few \citep{sch77}.

Radio observations identified many \ion{H}{2} regions \citep{van88},
indicating that the system has an on-going star forming activity at a star formation rate of $\sim$4~$M_{\odot}$~yr$^{-1}$ \citep{sco01}.
Two (and possibly another) core-collapse supernovae (SNe) recently occurred in the galaxy\footnote{See http://web.oapd.inaf.it/supern/cat/ for detail.}.

M\,51 has been observed many times in X-rays by the {\it{Einstein}} \citep{pal85}, {\it{ROSAT}} \citep{mar95,ehle95},
{\it{ASCA}} \citep{tera98,ptak99a}, {\it{BeppoSAX}} \citep{fuk01}, {\it{XMM-Newton}} \citep{dew05}, and {\it{Chandra}} \citep{liu02,tera04} observatories.
The number of X-ray sources increased from three by {\it{Einstein}} \citep{pal85} to 113 by {\it{Chandra}} \citep{tera04}, which includes nine ULXs \citep{dew05}.
We follow the nomenclature by \citet{tera04}.
Figure~\ref{fig1} (a) shows positions of the ULXs on the optical image by the Sloan Digital Sky Survey (SDSS) data.
All ULXs are distributed along galactic arms.
\citet{tera06} reported using the {\it{Hubble Space Telescope}} data
that four ULXs (sources-9, 37, 69, and 82) are located near the center or at the rim of star clusters.
They also identified one or more optical counterparts to six ULXs.

Previously published X-ray studies \citep[e.g.][]{dew05,tera06} were based on four X-ray data sets among seven.
We reduce all the available data including three unpublished sets.

\subsection{NGC\,4490/85}

NGC\,4490 and NGC\,4485, the former being a spiral and the latter being an irregular galaxy, are interacting with each other.
NGC\,4490/85 is one of the closest interacting galaxies as well as M\,51,
belonging to the CVn\,II group of galaxies \citep[][]{deVau76,via80,clem98,clem99,fri08}.
The size and the mass of NGC\,4490 are $\sim$15~kpc \citep{clem02} and $\sim$1.6$\times 10^{10}$~$M_{\odot}$ \citep{via80}, respectively,
while those of NGC\,4485 are $\sim$5.6~kpc \citep{clem02} and $\sim$2$\times 10^{9}$~$M_{\odot}$ \citep{clem99}, respectively.

Radio observations identified many \ion{H}{2} regions in NGC\,4490,
indicating an on-going star forming activity at a star formation rate of $\sim$4.7~$M_{\odot}$~yr$^{-1}$ \citep{clem99,clem02}.
Two core-collapse SNe recently occurred in the galaxy$^4$.

NGC\,4490/85 was observed several times with the {\it{ROSAT}} \citep{read97,rob00}, {\it{XMM-Newton}}, and {\it{Chandra}} observatories \citep{rob02,fri08}.
The number of known X-ray sources increased from five by {\it{ROSAT}} \citep{rob00} to 38 by {\it{Chandra}} \citep{fri08}, which include eight ULXs (ULX-1 to ULX-8).
We follow the nomenclature by \citet{fri08}.
Figure~\ref{fig2} (a) shows positions of the ULXs on the SDSS image.
ULX-2 belongs to NGC\,4485, while the others to NGC\,4490.
ULX-6 is close to but offset from the dynamical center of the galaxy \citep{rob02}, and the other ULXs in NGC\,4490 are distributed along galactic arms.
Six of them were observed by the {\it{Spitzer Space Telescope}}.
Five sources (ULX-2, 3, 4, 6, and 8) are likely to be an accreting X-ray source
based on the detection of some characteristic features of highly ionized species in the mid-infrared spectra,
while the remaining one (ULX-1) is more likely to be an SN remnant \citep{vaz07}.
Nevertheless, X-ray flux variation was found from ULX-1 \citep{fri08}, leading to the speculation that this source is a ULX associated with the SN remnant.

Previous X-ray studies \citep{rob02,fri08} presented long-term variations in flux and color in several ULXs and gave some phenomenological analysis.
We present long-term spectral variations of all the bright ULXs along with more physical models.

\begin{deluxetable*}{cccccccccccc}
\tabletypesize{\tiny}
\tablecaption{Source and Background Counts for M\,51.\label{table2}}
\tablewidth{0pt}
\tablecolumns{12}
\tablehead{
Name\tablenotemark{a} & \multicolumn{11}{c}{Counts\tablenotemark{b}} \\
\cline{2-12}
 & \multicolumn{3}{c}{{\it{Chandra}}} & & \multicolumn{7}{c}{{\it{XMM-Newton}}} \\
\cline{2-4} \cline{6-12}
 & C1 & C2 & C3 & & \multicolumn{3}{c}{X1} & & \multicolumn{3}{c}{X2} \\
\cline{6-8} \cline{10-12}
 & & & & & MOS-1 & MOS-2 & pn & & MOS-1 & MOS-2 & pn
}
\startdata
Source-5 & 213 (0.1) & 163 (0.2) & \phantom{1}397 (1.3) & & 321 (\phantom{1}59) & 352 (\phantom{1}44) & \phantom{1}917 (\phantom{1}102) & & \phantom{1}392 (\phantom{1}62) & \phantom{1}441 (\phantom{1}60) & \phantom{1}850 (\phantom{11}63) \\
Source-9 & 121 (0.1) & 195 (0.2) & \phantom{1}696 (1.2) & & 182 (\phantom{1}71) & 172 (\phantom{1}79) & \phantom{1}227 (\phantom{1}143) & & \phantom{1}281 (\phantom{1}75) & \phantom{1}328 (102) & \phantom{1}862 (\phantom{1}246) \\
Source-26 & \phantom{1}92 (0.8) & 229 (1.3) & \phantom{1}375 (3.2) & & 349 (410) & 456 (374) & 1148 (1212) & & \phantom{1}520 (512) & \phantom{1}535 (433) & 1271 (1151) \\
Source-37 & \phantom{11}2 (0.2) & 497 (0.3) & \phantom{111}3 (0.5) & & 263 (\phantom{1}97) & 141 (\phantom{1}87) & \phantom{1}283 (\phantom{1}244) & & \phantom{1}344 (139) & \phantom{1}431 (148) & \phantom{1}947 (\phantom{1}326) \\
Source-41 & 196 (0.3) & 352 (0.5) & \phantom{1}554 (2.3) & & 137 (\phantom{1}35) & 129 (\phantom{1}44) & \phantom{1}384 (\phantom{1}121) & & \phantom{1}254 (\phantom{1}49) & \phantom{1}245 (\phantom{1}62) & \phantom{1}576 (\phantom{1}163) \\
Source-63 & 105 (0.2) & 245 (0.3) & \phantom{1}184 (0.6) & & 187 (\phantom{1}81) & 222 (\phantom{1}78) & \phantom{1}448 (\phantom{1}217) & & \phantom{1}227 (123) & \phantom{1}262 (138) & \phantom{1}411 (\phantom{1}234)\\
{\bf Source-69} & 463 (0.9) & \phantom{1}42 (1.2) & 1698 (3.6) & & 302 (\phantom{1}97) & 309 (107) & \phantom{1}813 (\phantom{1}329) & & 1137 (152) & 1258 (200) & 2679 (\phantom{1}401) \\
{\bf Source-82} & 757 (0.3) & 778 (0.5) & 1791 (0.9) & & 513 (\phantom{1}76) & 514 (\phantom{1}51) & \phantom{1}465 (\phantom{11}95) & & \phantom{1}561 (\phantom{1}74) & \phantom{1}603 (\phantom{1}87) & \nodata\tablenotemark{c} \\
Source-12 & 191 (1.1) & 332 (2.3) & \phantom{1}466 (6.0) & & 195 (\phantom{1}37) & 178 (\phantom{1}24) & \phantom{1}462 (\phantom{1}114) & & \phantom{1}176 (101) & \phantom{1}121 (\phantom{1}82) & \phantom{1}321 (\phantom{1}126) \\
\tableline
 & \multicolumn{7}{c}{{\it{XMM-Newton}}} & & & & \\
\cline{2-8}
 & \multicolumn{3}{c}{X3} & & \multicolumn{3}{c}{X4} & & & & \\
\cline{2-4} \cline{6-8}
 & MOS-1 & MOS-2 & pn & & MOS-1 & MOS-2 & pn & & & & \\
\tableline
Source-5 & \phantom{1}509 (101) & \phantom{1}466 (101) & \phantom{1}411 (\phantom{1}88) & & \phantom{1}536 (\phantom{1}79) & \phantom{1}514 (\phantom{1}97) & \phantom{1}451 (\phantom{1}94) \\
Source-9 & \phantom{1}117 (119) & \phantom{1}130 (157) & \nodata\tablenotemark{c} & & \phantom{1}187 (\phantom{1}81) & \phantom{1}205 (\phantom{1}89) & \phantom{1}617 (236) \\
Source-26 & \phantom{1}607 (668) & \phantom{1}668 (596) & \nodata\tablenotemark{c} & & \phantom{1}420 (496) & \phantom{1}474 (441) & \phantom{1}376 (757) \\
Source-37 & \phantom{1}277 (162) & \phantom{1}265 (160) & \nodata\tablenotemark{c} & & \phantom{1}113 (113) & \phantom{1}108 (124) & \phantom{1}275 (252) \\
Source-41 & \phantom{1}319 (\phantom{1}95) & \phantom{1}265 (\phantom{1}82) & \phantom{1}691 (167) & & \phantom{1}212 (\phantom{1}43) & \phantom{1}229 (\phantom{1}57) & \phantom{1}593 (110) \\
Source-63 & \phantom{1}233 (141) & \phantom{1}217 (138) & \phantom{1}476 (193) & & \phantom{1}162 (\phantom{1}88) & \phantom{1}170 (101) & \phantom{1}366 (207) \\
{\bf Source-69} & 1114 (164) & 1306 (237) & 2887 (647) & & 1067 (173) & 1209 (207) & 2864 (561) \\
{\bf Source-82} & \phantom{1}653 (\phantom{1}87) & \phantom{1}577 (107) & 1448 (225) & & \phantom{1}461 (\phantom{1}70) & \phantom{1}458 (\phantom{1}78) & 1236 (194) \\
Source-12 & \phantom{1}157 (\phantom{1}67) & \phantom{1}129 (\phantom{1}66) & \phantom{1}374 (118) & & \phantom{1}118 (\phantom{1}50) & \phantom{1}130 (\phantom{1}38) & \phantom{1}146 (\phantom{1}91)
\enddata
\tablenotetext{a}{The nomenclatures follow \citet{dew05}. The two brightest ULXs, focused in this paper, are shown in the bold font.}
\tablenotetext{b}{Source counts (background counts). The background counts are normalized to the source extraction area.}
\tablenotetext{c}{The data are unavailable, because the source is located in a gap or a dead area of a chip.}
\end{deluxetable*}

\begin{deluxetable*}{cccccccc}
\tablecaption{Source and Background Counts for NGC\,4490/85.\label{table3}}
\tablewidth{0pt}
\tablecolumns{8}
\tablehead{
Name\tablenotemark{a} & \multicolumn{7}{c}{Counts\tablenotemark{b}} \\
\cline{2-8}
 & \multicolumn{3}{c}{{\it{Chandra}}} & & \multicolumn{3}{c}{{\it{XMM-Newton}}} \\
\cline{2-4} \cline{6-8}
 & C1$^{\prime}$ & C2$^{\prime}$ & C3$^{\prime}$ & & \multicolumn{3}{c}{X1$^{\prime}$} \\
\cline{6-8}
 & & & & & MOS-1 & MOS-2 & pn
}
\startdata
ULX-1 & \phantom{1}239 (0.2) & \phantom{1}424 (1.1) & \phantom{1}305 (0.7) & & \nodata & \nodata & \nodata \\
{\bf ULX-2} & 1243 (0.8) & 1772 (9.9) & 1436 (0.9) & & 609 (23.5) & 499 (21.0) & 403 (50.2) \\
{\bf ULX-3} & \phantom{1}566 (0.3) & 1148 (1.2) & 1209 (1.0) & & 339 (\phantom{1}4.8) & 309 (\phantom{1}4.2) & 554 (10.2) \\
{\bf ULX-4} & \phantom{1}455 (0.4) & 1291 (1.9) & 1375 (1.3) & & 365 (\phantom{1}4.8) & 419 (\phantom{1}4.2) & 775 (10.2) \\
ULX-5 & \phantom{11}30 (0.8) & \phantom{1}374 (1.7) & \phantom{1}561 (1.5) & & \nodata & \nodata & \nodata \\
{\bf ULX-6} & \phantom{1}473 (0.8) & \phantom{1}299 (1.9) & 1347 (2.3) & & 444 (\phantom{1}8.5) & 481 (\phantom{1}7.5) & 897 (18.1) \\
ULX-7 & \phantom{111}0 (0.3) & \phantom{111}1 (0.7) & \phantom{1}749 (0.9) & & \nodata & \nodata & \nodata \\
{\bf ULX-8} & \phantom{1}780 (0.4) & 1333 (1.2) & 1762 (1.4) & & 525 (23.5) & 484 (21.0) & 811 (50.2)
\enddata
\tablenotetext{a}{The nomenclatures follow \citet{fri08}. The five brightest ULXs, focused in this paper, are shown in the bold font.}
\tablenotetext{b}{Same with Table~\ref{table2}.}
\end{deluxetable*}

\begin{figure*}
\plotone{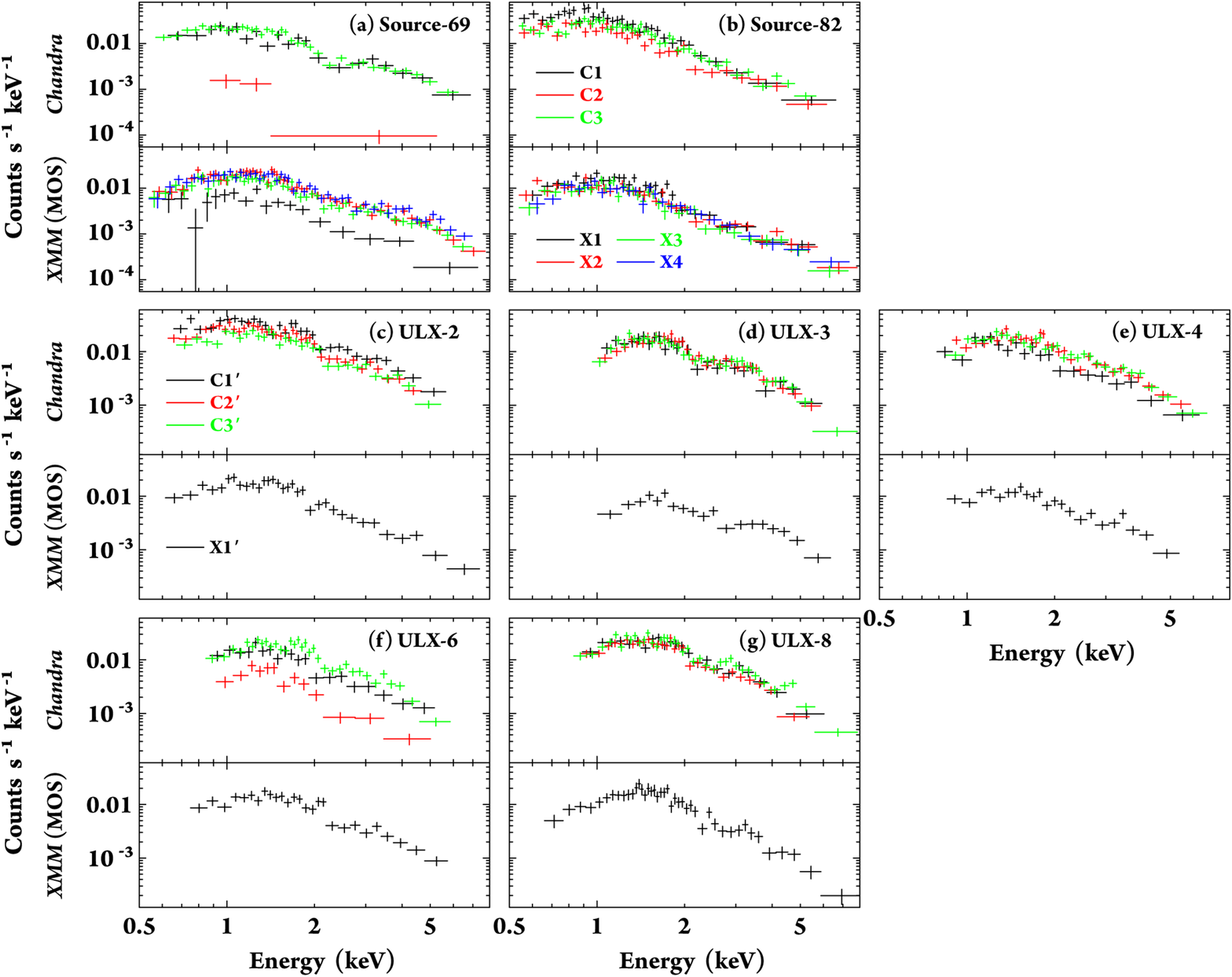}
\caption{Time-averaged spectra of the seven ULXs.
The top panels show the {\it{Chandra}} ACIS-S3 spectra by the three observations; C1, C1$^{\prime}$ ({\it{black}}), C2, C2$^{\prime}$ ({\it{red}}), and C3, C3$^{\prime}$ ({\it{green}}).
The bottom panels show the {\it{XMM-Newton}} merged MOS spectra by the four (M\,51) or one (NGC\,4490/85) observations;
X1, X1$^{\prime}$ ({\it{black}}), X2 ({\it{red}}), X3 ({\it{green}}), and X4 ({\it{blue}}).
\label{fig3}}
\end{figure*}

\section{Observations and Data Reduction}

We selected data sets with an exposure time longer than 10~ks for detailed spectral analysis.
Throughout this paper, we use X-ray events in the 0.5--8.0~keV energy band.
For M\,51, we analyzed three {\it{Chandra}} and four {\it{XMM-Newton}} archived data sets (Table~\ref{table1}).
We labeled the observations as C1-3 and X1-4, respectively.
C1, C2, C3, and X1 were used by \citet{liu02}, \citet{tera04}, \citet{dew05}, and \citet{tera06}.

For NGC\,4490/85, we reduced all the archived data sets of {\it{Chandra}} and {\it{XMM-Newton}} (Table~\ref{table1}).
The sets are the same with those used in \citet{fri08} and \citet{glad09}.
We labeled the {\it{Chandra}} and the {\it{XMM-Newton}} observations as C1$^{\prime}$-3$^{\prime}$ and X1$^{\prime}$, respectively.

\subsection{{\it{Chandra}}}

The {\it{Chandra X-ray Observatory}} \citep{wei02} has an unprecedented spatial resolution of $\sim$0\farcs5 at the optical axis.
The {\it{Chandra}} observations were conducted using the Advanced CCD for Imaging and Spectroscopy \citep[ACIS;][]{gar03},
which covers an energy range of 0.5--8.0~keV.
The data were taken with the full frame mode with the aim point in the S3 chip.
The field of view (FOV) of the chip is $\sim$500\arcsec$\times$500\arcsec, which contains the entire system of both M\,51 and NGC\,4490/85.

Using the {\it{Chandra}} Interactive Analysis of Observations (CIAO) version 4.0 and the ACIS Extract package \citep{bro02} version 2008-06-06,
we extracted the source and background events and constructed the energy spectra.
The source events were accumulated from a region around each source encircling 90\% of photons of a point-like source.
The background events were from an annulus around each source.

We list the number of counts for the ULXs identified in M\,51 \citep{dew05} and NGC\,4490/85 \citep{fri08} in Tables~\ref{table2} and \ref{table3}, respectively.
Among all the ULXs, we focus on the sources with a total count of more than 1000 at least in one observation,
which are practically bright enough for our spectral analysis.
Seven sources are thus selected; sources-69 and 82 for M\,51, and ULX-2, 3, 4, 6, and 8 for NGC\,4490/85.

\subsection{{\it{XMM-Newton}}}

The {\it{XMM-Newton}} \citep{jan01} observatory is equipped with the European Photon Imaging Camera (EPIC), which is comprised of three X-ray CCD cameras.
Two of the cameras are MOS arrays (MOS-1 and MOS-2) having an on-axis spatial resolution of $\sim$5\arcsec\ and sensitivity in the 0.15--12~keV band \citep{tur01},
while the remaining one is a pn array having an on-axis resolution of $\sim$6\arcsec\ and sensitivity at 0.15--15~keV \citep{stru01}.
Both instruments have a FOV with a radius of 30--40\arcmin, and cover the two studied systems entirely (Figures~\ref{fig1}b and \ref{fig2}b).
In the M\,51 observations, the thin (X1, X3, and X4) and medium (X2) filters were used.
All cameras were operated with the full frame mode.
In the NGC\,4490/85 observation (X1$^{\prime}$), the medium filter was used.
The two MOS and pn cameras were operated with the full frame mode and the extended full frame mode, respectively.

We used the Science Analysis System (SAS) version 7.1.0 for extracting events and generating response files.
Some high background time intervals were seen in all observations.
For three observations (M\,51 X2, X3, and X4), we excluded intervals with count rates larger than the average by more than 3$\sigma$.
For the remaining observations, we used all exposures because the effect is negligible with only a few counts in extracted source counts.

We focus on the seven sources selected in the {\it{Chandra}} data (Tables~\ref{table2} and \ref{table3}).
The source extraction regions are shown in Figures~\ref{fig1} (b) and \ref{fig2} (b).
We chose the source region with a radius of 9--20\arcsec\ to avoid contamination from other sources.
In M\,51, the background regions were extracted from an annulus around each source to offset the global diffuse emission.
In contrast, in NGC\,4490/85, a common background region devoid of bright sources (Figure~\ref{fig2}b) was adopted.
The background events have $\sim$100 counts for MOS and $\sim$300 counts for pn.

\section{Data Analysis and Results}

Hereafter, we call each exposure of a source as a ``sample''.
The total number of the samples is 34, since there are two sources with seven exposures for M\,51 and five sources with four exposures for NGC\,4490/85.

\begin{deluxetable*}{ccccccccccc}
\tabletypesize{\scriptsize}
\tablecaption{Best-fit Parameters for M\,51.\label{table4}}
\tablewidth{0pt}
\tablecolumns{11}

\tablehead{
Name & Data & Model\tablenotemark{a} & $N_{\rm{H}}^{\prime}$ & ${\it{\Gamma}}$ & $k T_{\rm{in}}$ & $R_{\rm{in}}$ & $M_{\rm{BH}}$ & $dM/dt$ & $L_{\rm{X}}$ & Red$-\chi^2$ \\
 & label &  & $(10^{22}\;\rm{cm}^{-2})$ &  & (keV) & (km) & $(M_{\odot})$ & $(L_{\rm{Edd}} \; c^{-2})$ & $(10^{38}\;\rm{ergs \; s^{-1}})$ & (d.o.f.)
}

\startdata

Source-69 & C1 & P & $ <0.08 $ & $ 1.1^{+0.2}_{-0.1} $ & \nodata & \nodata & \nodata & \nodata & $ 27 \pm 3 $ & 1.12(18)\phantom{1} \\ 
 &  & M & $ <0.03 $ & \nodata & $ 2.3^{+1.2}_{-0.4} $ & $ \phantom{1}26^{+26}_{-\phantom{1}9} $ & \nodata & \nodata & $ 18^{+\phantom{1}3}_{-\phantom{1}2} $ & 1.33(18)\phantom{1} \\ 
 &  & K & $ 0.15^{+0.07}_{-0.07} $ & \nodata & \nodata & \nodata & $ \phantom{1}5^{+\phantom{1}6}_{-\phantom{1}3} $ & $ 100^{+513}_{-\phantom{1}71} $ & $ 16^{+\phantom{1}1}_{-\phantom{1}3} $ & 1.99(18)\phantom{1} \\ 
\cline{2-11}
 & C2 & P & 0.06(fixed) & $ 3.1^{+1.1}_{-0.9} $ & \nodata & \nodata & \nodata & \nodata & $ <0.8 $ & 0.63(1)\phantom{11} \\ 
 &  & M & 0.06(fixed) & \nodata & $ 0.4^{+0.2}_{-0.1} $ & $ 123^{+143}_{-80} $ & \nodata & \nodata & $ <0.4 $ & 0.14(1)\phantom{11} \\ 
 &  & K & 0.06(fixed) & \nodata & \nodata & \nodata & $ 10^{+\phantom{1}3}_{-\phantom{1}6} $ & $ <2 $  & $ <0.6 $ & 0.18(1)\phantom{11} \\ 
\cline{2-11}
 & C3 & P & $ 0.05^{+0.04}_{-0.04} $ & $ 1.5^{+0.1}_{-0.1} $ & \nodata & \nodata & \nodata & \nodata & $ 29 \pm 2 $ & 0.80(34)\phantom{1} \\ 
 &  & M & $ <0.01 $ & \nodata & $ 1.5^{+0.2}_{-0.1} $ & $ \phantom{1}60^{+16}_{-10} $ & \nodata & \nodata & $ 16 \pm 1 $ & 1.94(34)\phantom{1} \\ 
 &  & K & $ 0.07^{+0.03}_{-0.03} $ & \nodata & \nodata & \nodata & $ \phantom{1}7^{+\phantom{1}4}_{-\phantom{1}3} $ & $ 100^{+335}_{-\phantom{1}62} $ & $ 19 \pm 1 $ & 1.13(34)\phantom{1} \\ 
\cline{2-11}
 & X1 & P & $ 0.14^{+0.11}_{-0.06} $ & $ 1.9^{+0.1}_{-0.2} $ & \nodata & \nodata & \nodata & \nodata & $ 12 \pm 1 $ & 0.86(40)\phantom{1} \\ 
 &  & M & $ <0.03 $ & \nodata & $ 1.2^{+0.2}_{-0.1} $ & $ \phantom{1}55^{+21}_{-13} $ & \nodata & \nodata & $ \phantom{1}6.1 \pm 0.6 $ & 1.19(40)\phantom{1} \\ 
 &  & K & $ 0.05^{+0.06}_{-0.05} $ & \nodata & \nodata & \nodata & $ <9 $ & $ >425 $ & $ \phantom{1}7.5 \pm 0.7 $ & 0.91(40)\phantom{1} \\ 
\cline{2-11}
 & X2 & P & $ 0.09^{+0.03}_{-0.03} $ & $ 1.7^{+0.0}_{-0.1} $ & \nodata & \nodata & \nodata & \nodata & $ 52 \pm 2 $ & 1.18(110) \\ 
 &  & M & $ 0.00 $ & \nodata & $ 1.5^{+0.1}_{-0.1} $ & $ \phantom{1}81^{+12}_{-\phantom{1}9} $ & \nodata & \nodata & $ 30 \pm 1 $ & 2.03(110) \\ 
 &  & K & $ 0.08^{+0.02}_{-0.02} $ & \nodata & \nodata & \nodata & $ 12^{+\phantom{1}2}_{-\phantom{1}5} $ & $ 100^{+254}_{-\phantom{1}41} $ & $ 34 \pm 1 $ & 1.33(110) \\ 
\cline{2-11}
 & X3 & P & $ 0.09^{+0.03}_{-0.03} $ & $ 1.6^{+0.1}_{-0.1} $ & \nodata & \nodata & \nodata & \nodata & $ 41 \pm 2 $ & 1.58(113) \\ 
 &  & M & $ 0.00 $ & \nodata & $ 1.7^{+0.1}_{-0.1} $ & $ \phantom{1}58^{+\phantom{1}8}_{-\phantom{1}9} $ & \nodata & \nodata & $ 25 \pm 1 $ & 2.23(113) \\ 
 &  & K & $ 0.10^{+0.02}_{-0.02} $ & \nodata & \nodata & \nodata & $ 10^{+\phantom{1}1}_{-\phantom{1}4} $ & $ 100^{+154}_{-\phantom{1}34} $ & $ 27^{+\phantom{1}2}_{-\phantom{1}1} $ & 1.76(113) \\ 
\cline{2-11}
 & X4 & P & $ 0.10^{+0.03}_{-0.03} $ & $ 1.5^{+0.1}_{-0.1} $ & \nodata & \nodata & \nodata & \nodata & $ 58 \pm 2 $ & 1.41(111) \\ 
 &  & M & $ <0.01 $ & \nodata & $ 1.8^{+0.2}_{-0.1} $ & $ \phantom{1}60^{+10}_{-\phantom{1}7} $ & \nodata & \nodata & $ 35 \pm 2 $ & 1.80(111) \\ 
 &  & K & $ 0.13^{+0.02}_{-0.02} $ & \nodata & \nodata & \nodata & $ 13^{+\phantom{1}2}_{-\phantom{1}5} $ & $ 100^{+238}_{-\phantom{1}30} $ & $ 38^{+\phantom{1}1}_{-\phantom{1}2} $ & 1.71(111) \\

\tableline

Source-82 & C1 & P & $ 0.11^{+0.06}_{-0.06} $ & $ 2.3^{+0.3}_{-0.2} $ & \nodata & \nodata & \nodata & \nodata & $ 28 \pm 3 $ & 0.84(30)\phantom{1} \\ 
 &  & M & $ <0.01 $ & \nodata & $ 0.7^{+0.1}_{-0.1} $ & $ 222^{+51}_{-47} $ & \nodata & \nodata & $ 13 \pm 1 $ & 1.48(30)\phantom{1} \\ 
 &  & K & $ 0.01^{+0.05}_{-0.01} $ & \nodata & \nodata & \nodata & $ 34^{+\phantom{1}7}_{-\phantom{1}9} $ & $ \phantom{11}7^{+\phantom{11}3}_{-\phantom{11}1} $ & $ 15^{+\phantom{1}2}_{-\phantom{1}1} $ & 0.95(30)\phantom{1} \\ 
\cline{2-11}
 & C2 & P & $ 0.02^{+0.03}_{-0.02} $ & $ 1.9^{+0.2}_{-0.2} $ & \nodata & \nodata & \nodata & \nodata & $ 16 \pm 1 $ & 1.13(30)\phantom{1} \\ 
 &  & M & $ <0.01 $ & \nodata & $ 0.9^{+0.1}_{-0.1} $ & $ 121^{+38}_{-36} $ & \nodata & \nodata & $ \phantom{1}8.5 \pm 0.9 $ & 2.56(30)\phantom{1} \\
 &  & K & $ <0.01 $ & \nodata & \nodata & \nodata & $ 12^{+\phantom{1}3}_{-\phantom{1}5} $ & $ \phantom{1}14^{+\phantom{1}23}_{-\phantom{11}5} $ & $ 11 \pm 1 $ & 1.31(30)\phantom{1} \\ 
\cline{2-11}
 & C3 & P & $ 0.12^{+0.05}_{-0.04} $ & $ 2.1^{+0.1}_{-0.1} $ & \nodata & \nodata & \nodata & \nodata & $ 25 \pm 2 $ & 1.60(34)\phantom{1} \\ 
 &  & M & $ <0.01 $ & \nodata & $ 0.9^{+0.1}_{-0.1} $ & $ 146^{+21}_{-20} $ & \nodata & \nodata & $ 12 \pm 1 $ & 2.27(34)\phantom{1} \\ 
 &  & K & $ 0.04^{+0.03}_{-0.04} $ & \nodata & \nodata & \nodata & $ 22^{+10}_{-\phantom{1}8} $ & $ \phantom{11}9^{+\phantom{11}7}_{-\phantom{11}2} $ & $ 14 \pm 1 $ & 1.72(34)\phantom{1} \\ 
\cline{2-11}
 & X1 & P & $ 0.21^{+0.07}_{-0.06} $ & $ 2.6^{+0.3}_{-0.2} $ & \nodata & \nodata & \nodata & \nodata & $ 27^{+\phantom{1}4}_{-\phantom{1}3} $ & 1.18(47)\phantom{1} \\ 
 &  & M & $ <0.03 $ & \nodata & $ 0.7^{+0.1}_{-0.1} $ & $ 191^{+39}_{-40} $ & \nodata & \nodata & $ 11 \pm 1 $ & 1.91(47)\phantom{1} \\ 
 &  & K & $ 0.08^{+0.05}_{-0.05} $ & \nodata & \nodata & \nodata & $ 37^{+\phantom{1}9}_{-\phantom{1}7} $ & $ \phantom{11}5^{+\phantom{11}1}_{-\phantom{11}1} $ & $ 13 \pm 1 $ & 1.45(47)\phantom{1} \\ 
\cline{2-11}
 & X2 & P & $ <0.04 $ & $ 1.8^{+0.2}_{-0.1} $ & \nodata & \nodata & \nodata & \nodata & $ 19 \pm 2 $ & 0.97(31)\phantom{1} \\ 
 &  & M & $ <0.01 $ & \nodata & $ 1.0^{+0.2}_{-0.1} $ & $ 107^{+34}_{-29} $ & \nodata & \nodata & $ 11 \pm 1 $ & 2.58(31)\phantom{1} \\ 
 &  & K & $ <0.01 $ & \nodata & \nodata & \nodata & $ 11^{+\phantom{1}2}_{-\phantom{1}5} $ & $ \phantom{1}20^{+\phantom{1}58}_{-\phantom{11}8} $ & $ 13 \pm 1 $ & 1.22(31)\phantom{1} \\ 
\cline{2-11}
 & X3 & P & $ 0.11^{+0.04}_{-0.04} $ & $ 2.2^{+0.2}_{-0.1} $ & \nodata & \nodata & \nodata & \nodata & $ 17 \pm 1 $ & 0.87(75)\phantom{1} \\ 
 &  & M & $ <0.01 $ & \nodata & $ 0.8^{+0.1}_{-0.1} $ & $ 152^{+26}_{-25} $ & \nodata & \nodata & $ \phantom{1}8.1 \pm 0.5 $ & 1.80(75)\phantom{1} \\ 
 &  & K & $ 0.01^{+0.03}_{-0.01} $ & \nodata & \nodata & \nodata & $ 17^{+\phantom{1}5}_{-\phantom{1}3} $ & $ \phantom{11}7^{+\phantom{11}4}_{-\phantom{11}2} $ & $ \phantom{1}9.6 \pm 0.6 $ & 1.13(75)\phantom{1} \\ 
\cline{2-11}
 & X4 & P & $ 0.14^{+0.05}_{-0.04} $ & $ 2.2^{+0.1}_{-0.2} $ & \nodata & \nodata & \nodata & \nodata & $ 20 \pm 2 $ & 0.83(62)\phantom{1} \\ 
 &  & M & $ <0.01 $ & \nodata & $ 0.9^{+0.1}_{-0.1} $ & $ 132^{+26}_{-25} $ & \nodata & \nodata & $ \phantom{1}9.6 \pm 0.6 $ & 1.47(62)\phantom{1} \\ 
 &  & K & $ 0.04^{+0.04}_{-0.04} $ & \nodata & \nodata & \nodata & $ 19^{+\phantom{1}8}_{-\phantom{1}7} $ & $ \phantom{11}8^{+\phantom{11}7}_{-\phantom{11}2} $ & $ 11 \pm 1 $ & 1.03(62)\phantom{1}

\enddata
\tablenotetext{a}{The abbreviations for the models: ``P'' for PL, ``M'' for MCD (the \texttt{diskbb} model in XSPEC), and ``K'' for Kawaguchi model.}
\end{deluxetable*}

\begin{deluxetable*}{ccccccccccc}
\tabletypesize{\scriptsize}
\tablecaption{Best-fit Parameters for NGC\,4490/85.\label{table5}}
\tablewidth{0pt}
\tablecolumns{11}

\tablehead{
Name & Data & Model\tablenotemark{a} & $N_{\rm{H}}^{\prime}$ & ${\it{\Gamma}}$ & $k T_{\rm{in}}$ & $R_{\rm{in}}$ & $M_{\rm{BH}}$ & $dM/dt$ & $L_{\rm{X}}$ & Red$-\chi^2$ \\
 & label &  & $(10^{22}\;\rm{cm}^{-2})$ &  & (keV) & (km) & $(M_{\odot})$ & $(L_{\rm{Edd}} \; c^{-2})$ & $(10^{38}\;\rm{ergs \; s^{-1}})$ & (d.o.f.)
}

\startdata

ULX-2 & C1$^{\prime}$ & P & $ 0.21^{+0.08}_{-0.07} $ & $ 1.6^{+0.2}_{-0.2} $ & \nodata & \nodata & \nodata & \nodata & $ 52 \pm 4 $ & 1.69(28) \\ 
 &  & M & $ 0.05^{+0.05}_{-0.05} $ & \nodata & $ 1.6^{+0.2}_{-0.2} $ & $ \phantom{1}73^{+22}_{-17} $ & \nodata & \nodata & $ 29 \pm 2 $ & 1.34(28) \\ 
 &  & K & $ 0.22^{+0.05}_{-0.04} $ & \nodata & \nodata & \nodata & $ 12^{+12}_{-\phantom{1}7} $ & $ 100^{+506}_{-74} $ & $ 35^{+\phantom{1}3}_{-\phantom{1}1} $ & 1.68(28) \\ 
\cline{2-11}
 & C2$^{\prime}$ & P & $ 0.21^{+0.06}_{-0.06} $ & $ 1.8^{+0.2}_{-0.2} $ & \nodata & \nodata & \nodata & \nodata & $ 40 \pm 2 $ & 0.94(35) \\ 
 &  & M & $ 0.05^{+0.04}_{-0.04} $ & \nodata & $ 1.3^{+0.2}_{-0.1} $ & $ \phantom{1}92^{+22}_{-18} $ & \nodata & \nodata & $ 20 \pm 1 $ & 0.77(35) \\ 
 &  & K & $ 0.18^{+0.04}_{-0.05} $ & \nodata & \nodata & \nodata & $ 18^{+10}_{-13} $ & $ >13 $ & $ 25^{+\phantom{1}3}_{-\phantom{1}2} $ & 0.87(35) \\ 
\cline{2-11}
 & C3$^{\prime}$ & P & $ 0.28^{+0.08}_{-0.04} $ & $ 1.8^{+0.1}_{-0.1} $ & \nodata & \nodata & \nodata & \nodata & $ 34 \pm 2 $ & 1.63(28) \\ 
 &  & M & $ 0.08^{+0.05}_{-0.05} $ & \nodata & $ 1.4^{+0.2}_{-0.1} $ & $ \phantom{1}72^{+18}_{-15} $ & \nodata & \nodata & $ 18 \pm 1 $ & 1.28(28) \\ 
 &  & K & $ 0.24^{+0.06}_{-0.07} $ & \nodata & \nodata & \nodata & $ 15^{+\phantom{1}7}_{-10} $ & $ >14 $ & $ 22 \pm 2 $ & 1.50(28) \\ 
\cline{2-11}
 & X1$^{\prime}$ & P & $ 0.27^{+0.07}_{-0.07} $ & $ 2.1^{+0.1}_{-0.2} $ & \nodata & \nodata & \nodata & \nodata & $ 40^{+\phantom{1}4}_{-\phantom{1}3} $ & 1.36(48) \\ 
 &  & M & $ 0.02^{+0.05}_{-0.02} $ & \nodata & $ 1.2^{+0.1}_{-0.1} $ & $ \phantom{1}97^{+21}_{-20} $ & \nodata & \nodata & $ 19 \pm 1 $ & 1.60(48) \\ 
 &  & K & $ 0.16^{+0.04}_{-0.04} $ & \nodata & \nodata & \nodata & $ 23^{+\phantom{1}9}_{-\phantom{1}6} $ & $ \phantom{1}16^{+10}_{-\phantom{1}4} $ & $ 24 \pm 2 $ & 1.40(48) \\

\tableline

ULX-3 & C1$^{\prime}$ & P & $ 0.82^{+0.35}_{-0.30} $ & $ 2.0^{+0.4}_{-0.3} $ & \nodata & \nodata & \nodata & \nodata & $ 36^{+12}_{-\phantom{1}7} $ & 1.05(22) \\ 
 &  & M & $ 0.37^{+0.24}_{-0.21} $ & \nodata & $ 1.5^{+0.4}_{-0.3} $ & $ \phantom{1}65^{+38}_{-25} $ & \nodata & \nodata & $ 18 \pm 2 $ & 1.14(22) \\ 
 &  & K & $ 0.65^{+0.21}_{-0.20} $ & \nodata & \nodata & \nodata & $ 17^{+16}_{-13} $ & $ >12 $ & $ 22 \pm 4 $ & 1.06(22) \\ 
\cline{2-11}
 & C2$^{\prime}$ & P & $ 1.39^{+0.25}_{-0.23} $ & $ 2.4^{+0.3}_{-0.3} $ & \nodata & \nodata & \nodata & \nodata & $ 58^{+21}_{-14} $ & 0.63(30) \\ 
 &  & M & $ 0.79^{+0.17}_{-0.15} $ & \nodata & $ 1.2^{+0.2}_{-0.1} $ & $ 113^{+30}_{-25} $ & \nodata & \nodata & $ 27 \pm 2 $ & 0.67(30) \\ 
 &  & K & $ 1.08^{+0.15}_{-0.15} $ & \nodata & \nodata & \nodata & $ 40^{+\phantom{1}8}_{-\phantom{1}7} $ & $ \phantom{1}14^{+\phantom{1}5}_{-\phantom{1}2} $ & $ 33^{+\phantom{1}5}_{-\phantom{1}6} $ & 0.60(30) \\ 
\cline{2-11}
 & C3$^{\prime}$ & P & $ 1.08^{+0.19}_{-0.17} $ & $ 2.1^{+0.2}_{-0.2} $ & \nodata & \nodata & \nodata & \nodata & $ 51^{+\phantom{1}9}_{-\phantom{1}7} $ & 0.90(33) \\ 
 &  & M & $ 0.60^{+0.13}_{-0.12} $ & \nodata & $ 1.4^{+0.2}_{-0.1} $ & $ \phantom{1}78^{+19}_{-16} $ & \nodata & \nodata & $ 23 \pm 2 $ & 0.84(33) \\ 
 &  & K & $ 0.89^{+0.12}_{-0.11} $ & \nodata & \nodata & \nodata & $ 29^{+\phantom{1}6}_{-10} $ & $ \phantom{1}17^{+11}_{-\phantom{1}4} $ & $ 29^{+\phantom{1}5}_{-\phantom{1}2} $ & 0.82(33) \\ 
\cline{2-11}
 & X1$^{\prime}$ & P & $ 0.85^{+0.22}_{-0.20} $ & $ 1.8^{+0.2}_{-0.1} $ & \nodata & \nodata & \nodata & \nodata & $ 44^{+\phantom{1}8}_{-\phantom{1}6} $ & 1.00(33) \\ 
 &  & M & $ 0.39^{+0.14}_{-0.13} $ & \nodata & $ 1.8^{+0.3}_{-0.2} $ & $ \phantom{1}50^{+17}_{-13} $ & \nodata & \nodata & $ 25 \pm 2 $ & 1.10(33) \\ 
 &  & K & $ 0.74^{+0.15}_{-0.10} $ & \nodata & \nodata & \nodata & $ 12^{+15}_{-\phantom{1}7} $ & $ >23 $ & $ 29^{+\phantom{1}4}_{-\phantom{1}3} $ & 1.00(33) \\

\tableline

ULX-4 & C1$^{\prime}$ & P & $ 0.49^{+0.20}_{-0.16} $ & $ 1.9^{+0.3}_{-0.2} $ & \nodata & \nodata & \nodata & \nodata & $ 25^{+\phantom{1}4}_{-\phantom{1}3} $ & 1.08(17) \\ 
 &  & M & $ 0.22^{+0.13}_{-0.12} $ & \nodata & $ 1.5^{+0.4}_{-0.3} $ & $ \phantom{1}55^{+29}_{-20} $ & \nodata & \nodata & $ 13^{+\phantom{1}2}_{-\phantom{1}1} $ & 1.15(17) \\ 
 &  & K & $ 0.41^{+0.13}_{-0.12} $ & \nodata & \nodata & \nodata & $ 12^{+\phantom{1}6}_{-\phantom{1}9} $ & $ >12 $ & $ 16^{+\phantom{1}2}_{-\phantom{1}3} $ & 1.08(17) \\ 
\cline{2-11}
 & C2$^{\prime}$ & P & $ 0.63^{+0.17}_{-0.15} $ & $ 1.9^{+0.1}_{-0.2} $ & \nodata & \nodata & \nodata & \nodata & $ 51^{+\phantom{1}7}_{-\phantom{1}6} $ & 1.36(29) \\ 
 &  & M & $ 0.29^{+0.11}_{-0.10} $ & \nodata & $ 1.5^{+0.2}_{-0.2} $ & $ \phantom{1}81^{+24}_{-19} $ & \nodata & \nodata & $ 27 \pm 2 $ & 1.24(29) \\ 
 &  & K & $ 0.54^{+0.11}_{-0.11} $ & \nodata & \nodata & \nodata & $ 26^{+\phantom{1}9}_{-20} $ & $ >16 $ & $ 33^{+\phantom{1}5}_{-\phantom{1}2} $ & 1.29(29) \\ 
\cline{2-11}
 & C3$^{\prime}$ & P & $ 0.64^{+0.13}_{-0.06} $ & $ 1.9^{+0.1}_{-0.1} $ & \nodata & \nodata & \nodata & \nodata & $ 42 \pm 4 $ & 1.63(28) \\ 
 &  & M & $ 0.34^{+0.09}_{-0.08} $ & \nodata & $ 1.5^{+0.2}_{-0.2} $ & $ \phantom{1}70^{+17}_{-14} $ & \nodata & \nodata & $ 22 \pm 1 $ & 1.42(28) \\ 
 &  & K & $ 0.57^{+0.09}_{-0.11} $ & \nodata & \nodata & \nodata & $ 19^{+\phantom{1}9}_{-14} $ & $ >16 $ & $ 27 \pm 2 $ & 1.53(28) \\ 
\cline{2-11}
 & X1$^{\prime}$ & P & $ 0.60^{+0.12}_{-0.11} $ & $ 2.0^{+0.2}_{-0.1} $ & \nodata & \nodata & \nodata & \nodata & $ 50^{+\phantom{1}7}_{-\phantom{1}5} $ & 1.32(44) \\ 
 &  & M & $ 0.30^{+0.08}_{-0.07} $ & \nodata & $ 1.3^{+0.1}_{-0.1} $ & $ \phantom{1}99^{+20}_{-17} $ & \nodata & \nodata & $ 26 \pm 2 $ & 0.98(44) \\ 
 &  & K & $ 0.50^{+0.08}_{-0.07} $ & \nodata & \nodata & \nodata & $ 34^{+\phantom{1}5}_{-\phantom{1}8} $ & $ \phantom{1}17^{+\phantom{1}7}_{-\phantom{1}3} $ & $ 31^{+\phantom{1}3}_{-\phantom{1}2} $ & 1.14(44) \\

\tableline

ULX-6 & C1$^{\prime}$ & P & $ 0.35^{+0.23}_{-0.18} $ & $ 1.8^{+0.2}_{-0.3} $ & \nodata & \nodata & \nodata & \nodata & $ 23^{+\phantom{1}4}_{-\phantom{1}3} $ & 0.72(17) \\ 
 &  & M & $ 0.10^{+0.15}_{-0.10} $ & \nodata & $ 1.4^{+0.5}_{-0.3} $ & $ \phantom{1}53^{+37}_{-22} $ & \nodata & \nodata & $ 12^{+\phantom{1}2}_{-\phantom{1}1} $ & 0.79(17) \\ 
 &  & K & $ 0.28^{+0.15}_{-0.12} $ & \nodata & \nodata & \nodata & $ 10^{+\phantom{1}7}_{-\phantom{1}8} $ & $ >11 $ & $ 15 \pm 2 $ & 0.73(17) \\ 
\cline{2-11}
 & C2$^{\prime}$ & P & $ 0.60^{+0.35}_{-0.20} $ & $ 2.6^{+0.6}_{-0.4} $ & \nodata & \nodata & \nodata & \nodata & $ \phantom{1}7^{+\phantom{1}7}_{-\phantom{1}3} $ & 1.09(10) \\ 
 &  & M & $ 0.15^{+0.22}_{-0.15} $ & \nodata & $ 0.9^{+0.3}_{-0.2} $ & $ \phantom{1}64^{+41}_{-29} $ & \nodata & \nodata & $ \phantom{1}3.0^{+0.6}_{-0.5} $ & 1.45(10) \\ 
 &  & K & $ 0.35^{+0.23}_{-0.21} $ & \nodata & \nodata & \nodata & $ 12^{+\phantom{1}7}_{-\phantom{1}6} $ & $ \phantom{1}15^{+\phantom{1}4}_{-\phantom{1}1} $ & $ \phantom{1}3.8^{+1.3}_{-1.0} $ & 1.24(10) \\ 
\cline{2-11}
 & C3$^{\prime}$ & P & $ 0.88^{+0.16}_{-0.14} $ & $ 2.4^{+0.2}_{-0.2} $ & \nodata & \nodata & \nodata & \nodata & $ 53^{+10}_{-\phantom{1}7} $ & 1.49(31) \\ 
 &  & M & $ 0.45^{+0.10}_{-0.09} $ & \nodata & $ 1.1^{+0.1}_{-0.1} $ & $ 124^{+24}_{-21} $ & \nodata & \nodata & $ 21^{+\phantom{1}2}_{-\phantom{1}1} $ & 0.93(31) \\ 
 &  & K & $ 0.65^{+0.10}_{-0.09} $ & \nodata & \nodata & \nodata & $37^{+\phantom{1}5}_{-\phantom{1}4} $ & $ \phantom{1}11^{+\phantom{1}2}_{-\phantom{1}2} $ & $ 26 \pm 2 $ & 1.06(31) \\ 
\cline{2-11}
 & X1$^{\prime}$ & P & $ 0.41^{+0.08}_{-0.08} $ & $ 2.1^{+0.1}_{-0.1} $ & \nodata & \nodata & \nodata & \nodata & $ 44 \pm 4 $ & 1.12(52) \\ 
 &  & M & $ 0.11^{+0.05}_{-0.05} $ & \nodata & $ 1.4^{+0.1}_{-0.1} $ & $ \phantom{1}86^{+16}_{-14} $ & \nodata & \nodata & $ 23 \pm 1 $ & 1.07(52) \\ 
 &  & K & $ 0.30^{+0.05}_{-0.05} $ & \nodata & \nodata & \nodata & $ 27^{+\phantom{1}7}_{-\phantom{1}7} $ & $ \phantom{1}17^{+\phantom{1}8}_{-\phantom{1}4} $ & $ 27^{+\phantom{1}2}_{-\phantom{1}3} $ & 1.02(52) \\

\tableline

ULX-8 & C1$^{\prime}$ & P & $ 0.86^{+0.24}_{-0.20} $ & $ 2.3^{+0.3}_{-0.3} $ & \nodata & \nodata & \nodata & \nodata & $ 56^{+15}_{-\phantom{1}9} $ & 1.45(20) \\ 
 &  & M & $ 0.46^{+0.15}_{-0.13} $ & \nodata & $ 1.2^{+0.2}_{-0.1} $ & $ 116^{+34}_{-28} $ & \nodata & \nodata & $ 24 \pm 2 $ & 0.96(20) \\ 
 &  & K & $ 0.68^{+0.15}_{-0.14} $ & \nodata & \nodata & \nodata & $ 37^{+\phantom{1}8}_{-\phantom{1}8} $ & $ \phantom{1}12^{+\phantom{1}5}_{-\phantom{1}2} $ & $ 30 \pm 4 $ & 1.13(20) \\ 
\cline{2-11}
 & C2$^{\prime}$ & P & $ 0.90^{+0.16}_{-0.14} $ & $ 2.6^{+0.2}_{-0.2} $ & \nodata & \nodata & \nodata & \nodata & $ 46^{+13}_{-\phantom{1}9} $ & 1.28(26) \\ 
 &  & M & $ 0.43^{+0.10}_{-0.09} $ & \nodata & $ 1.0^{+0.1}_{-0.1} $ & $ 147^{+28}_{-25} $ & \nodata & \nodata & $ 19 \pm 2 $ & 0.75(26) \\ 
 &  & K & $ 0.60^{+0.12}_{-0.09} $ & \nodata & \nodata & \nodata & $ 40^{+13}_{-\phantom{1}5} $ & $ \phantom{11}9^{+\phantom{1}1}_{-\phantom{1}2} $ & $ 23^{+\phantom{1}3}_{-\phantom{1}2} $ & 0.82(26) \\ 
\cline{2-11}
 & C3$^{\prime}$ & P & $ 0.74^{+0.12}_{-0.11} $ & $ 2.0^{+0.2}_{-0.1} $ & \nodata & \nodata & \nodata & \nodata & $ 59^{+\phantom{1}7}_{-\phantom{1}5} $ & 1.52(36) \\ 
 &  & M & $ 0.38^{+0.08}_{-0.07} $ & \nodata & $ 1.4^{+0.1}_{-0.1} $ & $ \phantom{1}90^{+18}_{-16} $ & \nodata & \nodata & $ 28 \pm 2 $ & 1.47(36) \\ 
 &  & K & $ 0.61^{+0.08}_{-0.07} $ & \nodata & \nodata & \nodata & $ 32^{+\phantom{1}4}_{-10} $ & $ \phantom{1}20^{+12}_{-\phantom{1}4} $ & $ 35 \pm 4 $ & 1.41(36) \\ 
\cline{2-11}
 & X1$^{\prime}$ & P & $ 0.64^{+0.10}_{-0.09} $ & $ 2.5^{+0.2}_{-0.2} $ & \nodata & \nodata & \nodata & \nodata & $ 54^{+\phantom{1}9}_{-\phantom{1}7} $ & 0.88(58) \\ 
 &  & M & $ 0.26^{+0.07}_{-0.06} $ & \nodata & $ 1.0^{+0.1}_{-0.1} $ & $ 152^{+28}_{-25} $ & \nodata & \nodata & $ 21 \pm 1 $ & 0.92(58) \\ 
 &  & K & $ 0.39^{+0.08}_{-0.06} $ & \nodata & \nodata & \nodata & $ 38^{+10}_{-\phantom{1}3} $ & $ \phantom{1}10^{+\phantom{1}1}_{-\phantom{1}3} $ & $ 26^{+\phantom{1}2}_{-\phantom{1}3} $ & 0.86(58)

\enddata
\tablenotetext{a}{Same with Table~\ref{table4}.}
\end{deluxetable*}

\subsection{Light Curves}

Using the X-ray timing analysis package XRONOS version 5.2.1, we created the background-subtracted light curves of the seven ULXs.
We binned the curves with 1000~s~bin$^{-1}$ and fitted them with a constant flux model.
For the {\it{XMM-Newton}} observations, 
we considered that the flux is variable when neither of the MOS and pn light curves are fitted with a constant model.

As a result, we found significant variations (95\% significance) only from one source (M\,51 source-69).
This source showed a significant short-term variation in most samples (C1, C3, X2, X3, and X4).
We confirmed the periodicity of $\sim$7000~s in the flux in C1 as was reported by \citet{liu02}.

\begin{figure*}
\plotone{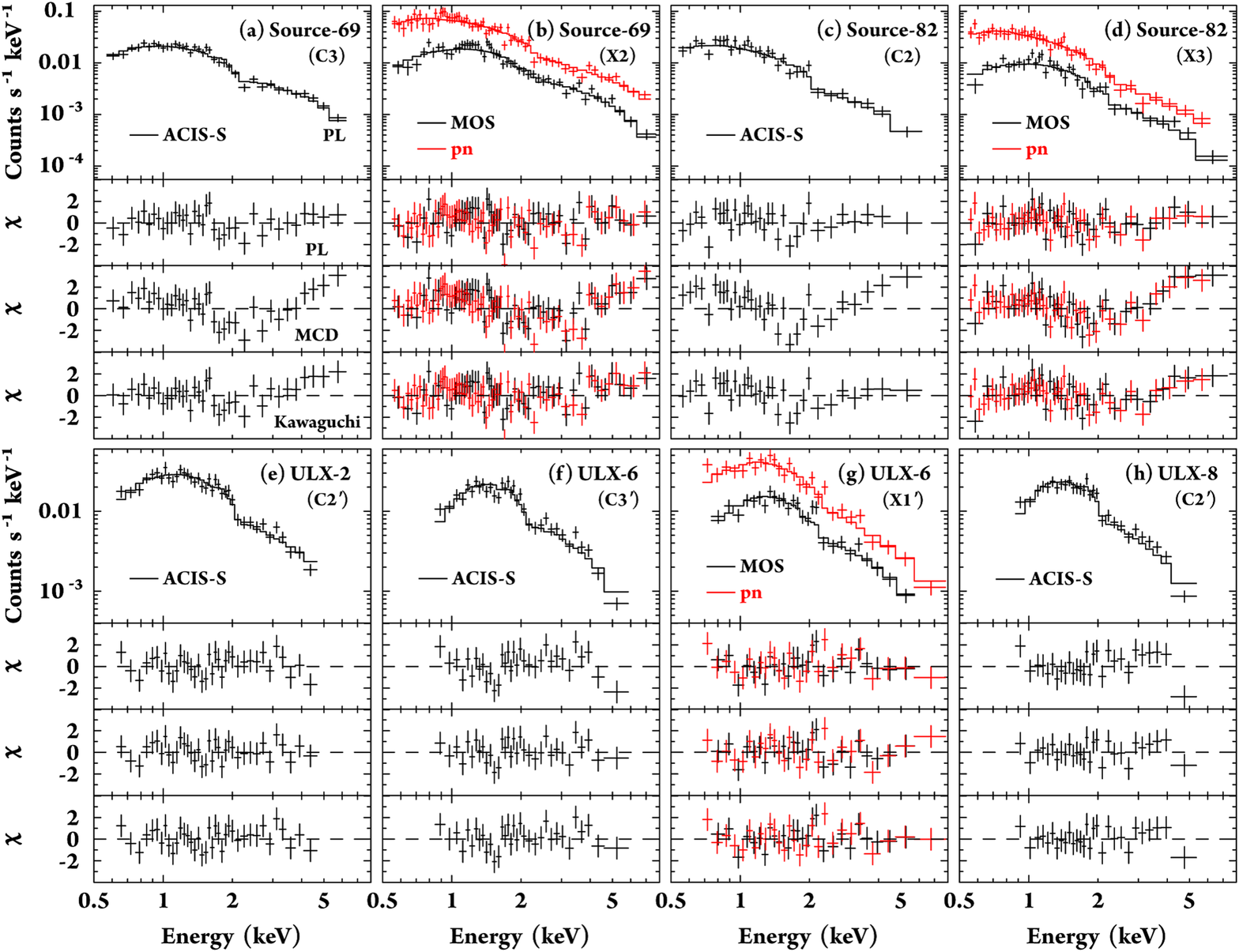}
\caption{Examples of spectral fitting in M\,51 (a--d) and NGC\,4490/85 (e--h).
Spectra are shown for different sources with different instruments as indicated by the labels and colors.
The top panels show the spectra (\textit{pluses}) and the best-fit PL model (\textit{solid histogram}),
and the other panels show the residuals from the best-fit by a PL, MCD, and Kawaguchi model.
\label{fig4}}
\end{figure*}

\subsection{Spectra}

Figure~\ref{fig3} shows the background-subtracted spectra of all the samples.
All were binned with at least 20~counts~bin$^{-1}$.
All spectra are featureless.
Some sources show flux variation among different observations.

\subsubsection{Fitting Models}

We used the X-ray spectral fitting package XSPEC version 11.3.2 for the spectral analysis.
For the interstellar extinction, we used the {\texttt{wabs}} model \citep{mor83};
for the hydrogen column densities,
we assumed the fixed Galactic extinction of 1.57$\times 10^{20}$~cm$^{-2}$ for M\,51 or 1.78$\times 10^{20}$~cm$^{-2}$ for NGC\,4490/85 \citep{dic90}
and a thawed additional extinction for each source.
For the {\it{XMM-Newton}} EPIC data, we fitted the merged MOS and pn spectra simultaneously.

We first applied a PL model and an MCD model \citep[\texttt{diskbb} in XSPEC;][]{mit84},
which are commonly used to fit the continuum emission of Galactic BHBs and ULXs.
The MCD model has the innermost disk temperature $T_{\rm{in}}$ as one of the parameters.
Using the relation $L_{\rm{bol}}$~$=$~$4 \pi (R_{\rm{in}}/\xi)^2 \sigma (T_{\rm{in}}/\kappa)^4$,
we can calculate the innermost disk radius $R_{\rm{in}}$ \citep{max00}.
Here $L_{\rm{bol}}$ is the bolometric luminosity, $\sigma$ is the Stefan-Boltzmann constant,
$\kappa$~$\sim$~1.7 is the spectral hardening factor \citep{sim95}, which is the ratio of the color temperature to the effective temperature,
and $\xi$~$=$~0.412 is the correction factor for the inner boundary condition \citep{kubo98}.
For non-spinning BHs, the BH mass $M_{\rm{BH}}$ can be estimated from $R_{\rm{in}}$ using the relation $R_{\rm{in}}$~$=$~$8.86(M_{\rm{BH}}/M_{\odot})$~km \citep{max00}.
For the MCD model with $k T_{\rm{in}}$~=~0.7--2.0~keV, the 0.5-8.0~keV band contains $>$80\% of the entire spectrum,
thus we approximate the X-ray luminosity in this band as the bolometric luminosity.

We also fitted all the spectra with a slim disk model by \citet{kawa03}\footnote{See http://heasarc.nasa.gov/xanadu/xspec/models/slimdisk.html for detail.}.
The ``Kawaguchi model'' takes into account of the Comptonization and the relativistic effects, and can constrain the BH mass and mass accretion rate as model parameters.
We fixed the viscous parameter $\alpha$ to be 0.1 \citep{kiki06}.
This model was successfully applied to some bright ULXs,
including M\,33 X-8 \citep{fos06}, NGC\,1313 X-2, NGC\,4559 X-7, X-10, NGC\,5204 X-1 \citep{kiki06}, and M\,82 X-1 \citep{oka06}.

\begin{figure}
\plotone{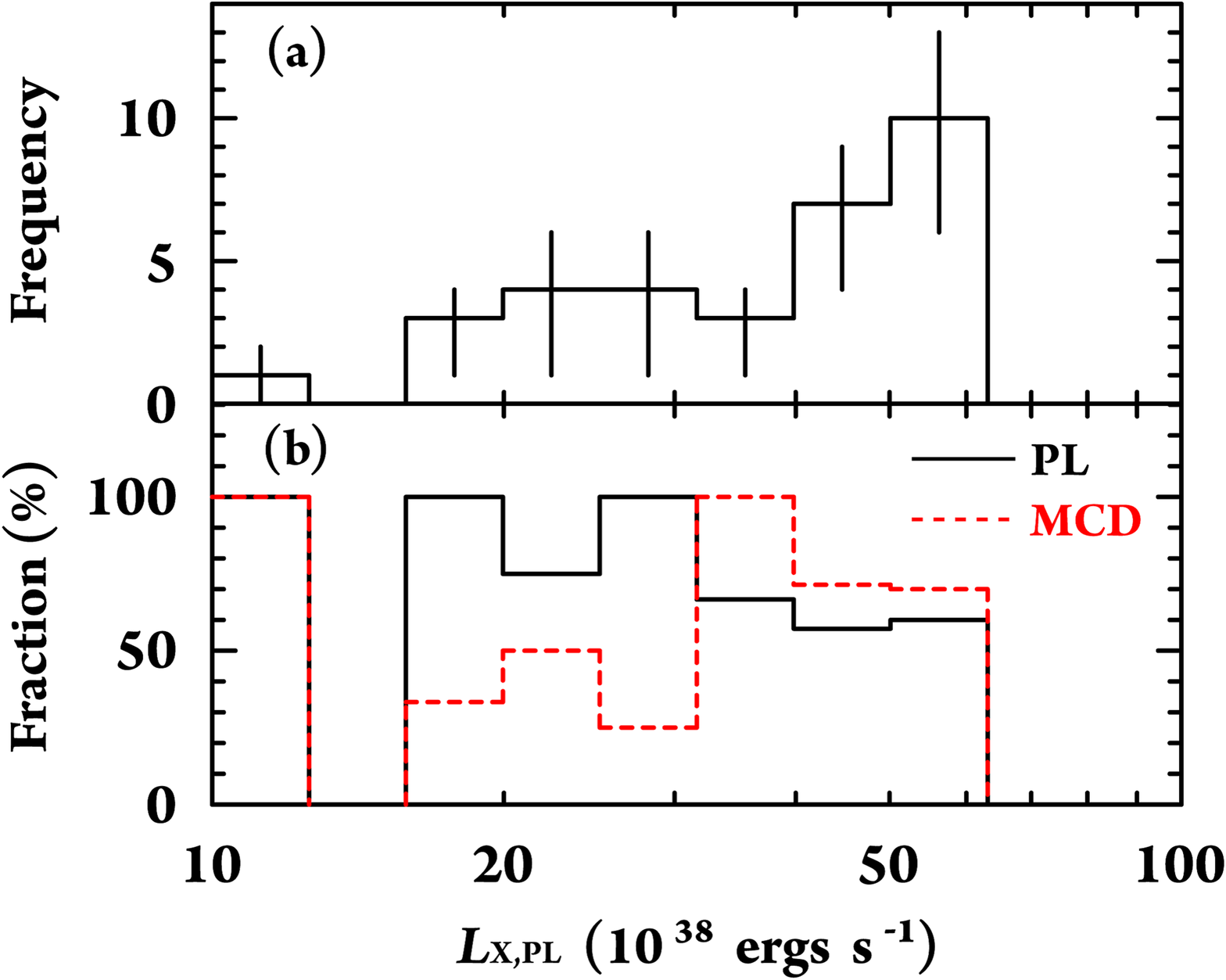}
\caption{Luminosity distribution of (a) the number of samples and
(b) the fraction of successful fits by the PL ({\it{black}}) and the MCD ({\it{red}}) models.
The 1$\sigma$ uncertainty, which is calculated from the Poisson distribution, is given in (a).
\label{fig5}}
\end{figure}

\subsubsection{Fitting Results}

We summarize all the best-fit parameters in Tables~\ref{table4} and \ref{table5} for samples in M\,51 and NGC\,4490/85, respectively.
$N_{\rm{H}}^{\prime}$, ${\it{\Gamma}}$, and $dM/dt$ indicate
the absorption column density additional to that in our Galaxy, the photon index, and the mass accretion rate, respectively.
Using the absorption-corrected flux
($f_{\rm{X,PL}}^{\prime}$, $f_{\rm{X,MCD}}^{\prime}$, and $f_{\rm{X,Kaw}}^{\prime}$ respectively for the PL, MCD, and Kawaguchi models),
the luminosity is calculated as $L_{\rm{X,PL}}$~$=$~$4\pi D^2 f_{\rm{X,PL}}^{\prime}$,
$L_{\rm{X,MCD}}$~$=$~$2\pi D^2 f_{\rm{X,MCD}}^{\prime} (\cos{i})^{-1}$, and $L_{\rm{X,Kaw}}$~$=$~$2\pi D^2 f_{\rm{X,Kaw}}^{\prime} (\cos{i})^{-1}$.
Here, $D$ is the distance to each galaxy, and $i$ is the inclination of the disk.
We assumed isotropic emission for the PL, and a moderate inclination for the MCD and Kawaguchi models ($i$~$=$~45$^{\circ}$).
Errors of all the parameters are calculated at the 90\% confidence level.
Figure~\ref{fig4} shows examples of the spectra and the best-fit models.

Although short-term flux and spectral hardness variation are found in M\,51 source-69 in C1,
we averaged each exposure because the paucity of the counts do not allow us to conduct time-sliced spectroscopy.
This source was very faint in C2, and $N_{\rm{H}}^{\prime}$ is fixed at 6.0$\times$10$^{20}$~cm$^{-1}$,
which is the median value obtained from the other six observations.

\begin{figure*}
\plotone{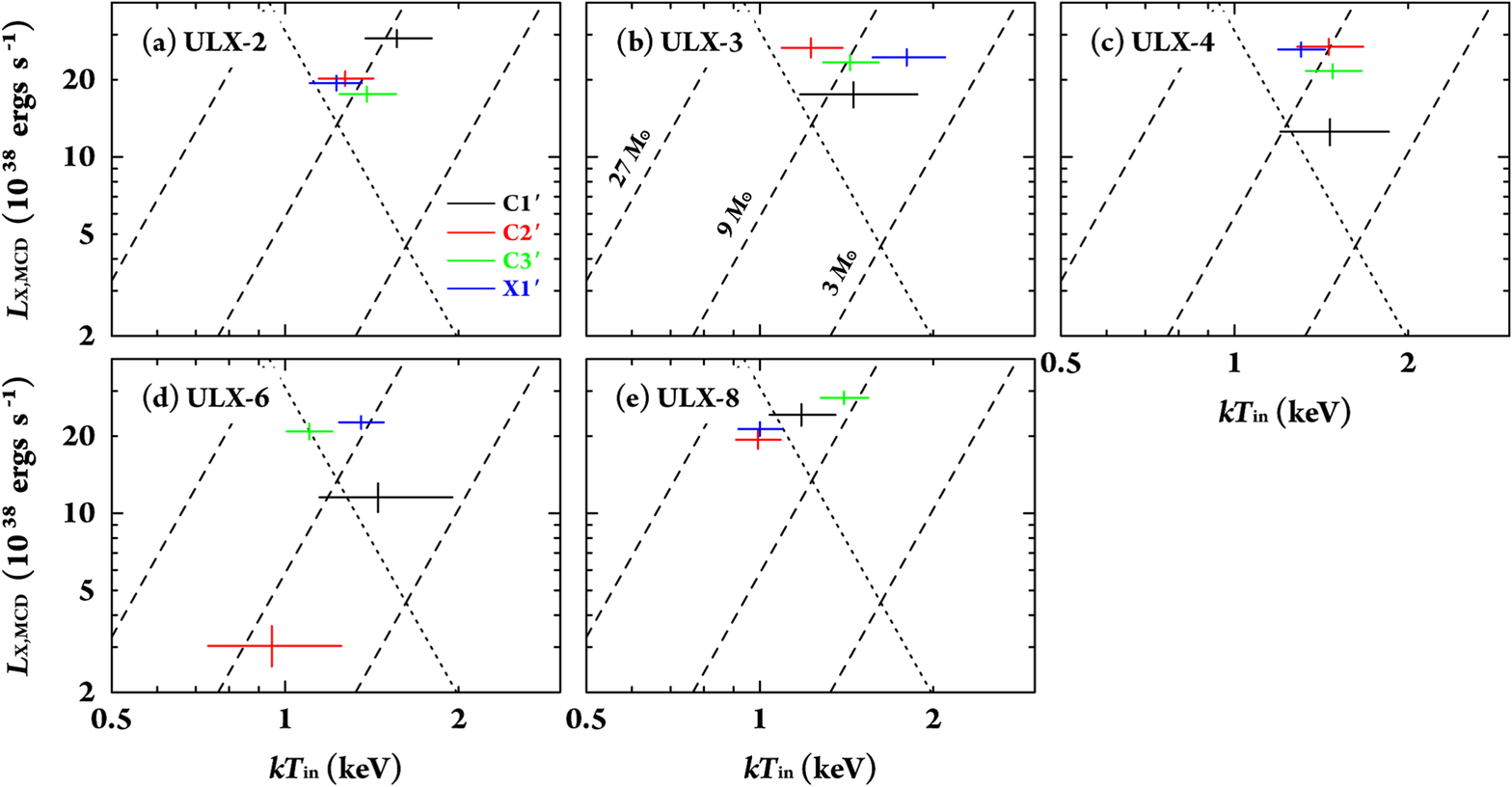}
\caption{Plot of X-ray luminosity $L_{\rm{X,MCD}}$ against the innermost disk temperature $k T_{\rm{in}}$ based on the MCD fitting results.
The results of different observations are shown in different colors.
The dashed lines show the $L_{\rm{X,MCD}}$~$\propto$~${T_{\rm{in}}}^4$ relation with several representative masses,
while the dotted line indicates the Eddington luminosity for the standard disk.
\label{fig6}}
\end{figure*}

\section{Discussion}

\subsection{ULX Spectral States and Transitions}

Based on the spectral fitting results presented in $\S$~4, we investigate the spectral states of ULXs and their transitions.
It is desirable that the states be defined for individual sources.
However, given the low photon counts of each source, starting with individual details may lead to an unclear view.
We thus first investigate the overall trend of all samples to constrain states of ULXs as a whole.
In this section, we develop a discussion under the working hypothesis
that all ULXs have a similar mass and have the same states and follow a similar pattern of state transitions. 
The working hypothesis is checked with individual sources in $\S$~5.2.

We constructed a histogram of the samples fitted successfully with the PL or the MCD model 
that represents PL-like and curved spectra, respectively (Figure~\ref{fig5}).
We consider that the fitting with a null hypothesis probability of $>$5\% is successful.
The best-fit luminosities derived by the PL fitting are used, because the derived luminosities hardly depend on the fitted models.

In the plot, it is noticeable that
the fraction of successful fits by the MCD model is higher within the range of 3--6$\times$10$^{39}$~ergs~s$^{-1}$ for 20~samples
and that by the PL model is higher within the range of 1.5--3$\times$10$^{39}$~ergs~s$^{-1}$ for 11~samples (Figure~\ref{fig5}b).
In order to test the claim statistically,
we simulated numerous spectra containing the two components (PL and MCD) with randomly chosen parameters with varying contrasts,
fitted them with one component model of either the PL or the MCD, and derived the fraction of the successful fits.
As a result, we found no trend that the MCD model is favored toward brighter luminosity.
Indeed, the PL model was favored.
Thus, we consider that the higher fraction of successful MCD fits toward the brighter luminosity is not a statistical artifact.

The bimodal structure in the luminosity is unlikely to be explained by the bimodal distribution of the BH mass or the inclination of the system.
This is because the distribution of the mass or the angle should be continuous, thus such a bimodal distribution is not natural.
Rather, we speculate that there are two distinctive states corresponding to each luminosity range.
We hereafter call them the ``{\it{bright}}'' state and the ``{\it{faint}}'' state.
The two states match the ``soft'' and the ``hard'' states of the ULXs in \citet{kubo01}.
Based on the number of samples in each state, the frequency ratio {\it{bright}}:{\it{faint}} is about 2:1.
The transition between the two occurs at $\sim$3$\times$10$^{39}$~ergs~s$^{-1}$ (hereafter the ``border luminosity'').
\citet{glad09} also claimed that the changes in spectral state occur
at a similar luminosity of $\sim$2$\times$10$^{39}$~ergs~s$^{-1}$ using samples in NGC\,4490/85.

Some samples do not belong to either of the two states.
Three samples (M\,51 source-69 in C2, X1, and NGC\,4490/85 ULX-6 in C2$^{\prime}$) show luminosities fainter than the tail of the fainter range.
These spectra may belong to the third state, which we call the ``{\it{dim}}'' state.
We cannot constrain its representative spectral model for the paucity of statistics.

Two types of luminosity variation are recognized.
We designate the variations within one state as the ``{\it{intra-state}}'' variations,
and those across two or more states as the ``{\it{inter-state}}'' transitions.

\begin{figure*}
\plotone{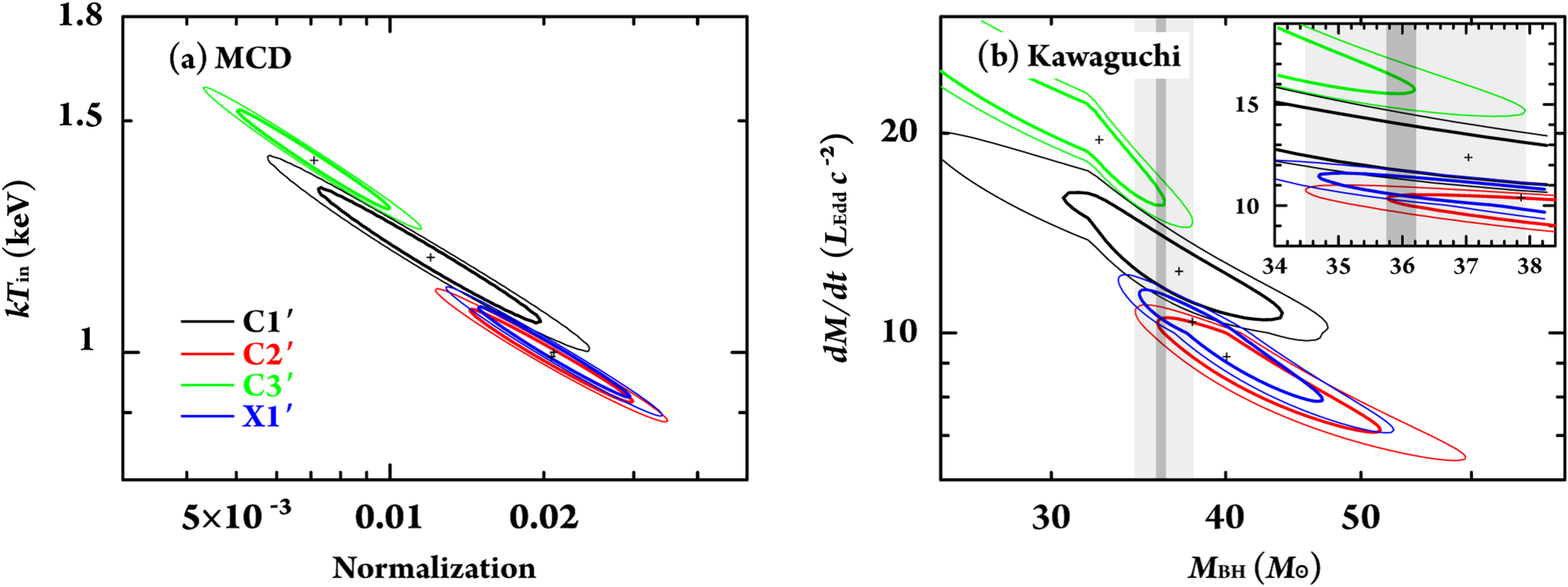}
\caption{Significance contours of the best-fit physical parameters by two different models for NGC\,4490/85 ULX-8;
(a) $k T_{\rm{in}}$ versus the normalization ($\propto$${M_{\rm{BH}}}^2$) by the MCD model.
(b) $dM/dt$ versus $M_{\rm{BH}}$ by the Kawaguchi model.
The results in different observations are shown with different colors: 
C1$^{\prime}$ ({\it black}), C2$^{\prime}$ ({\it red}), C3$^{\prime}$ ({\it green}), and X1$^{\prime}$ ({\it blue}).
The two confidence levels are shown for 68\% ({\it {thick solid}}) and 90\% ({\it {thin solid}}) ranges for the two parameters.
The inset at the top right in (b) is a close-up view of the center in the linear scale.
The dark and light gray shaded ranges indicate the common $M_{\rm{BH}}$ ranges among the four samples with a 68\% and 90\% confidence, respectively.
\label{fig7}}
\end{figure*}

\begin{figure}
\plotone{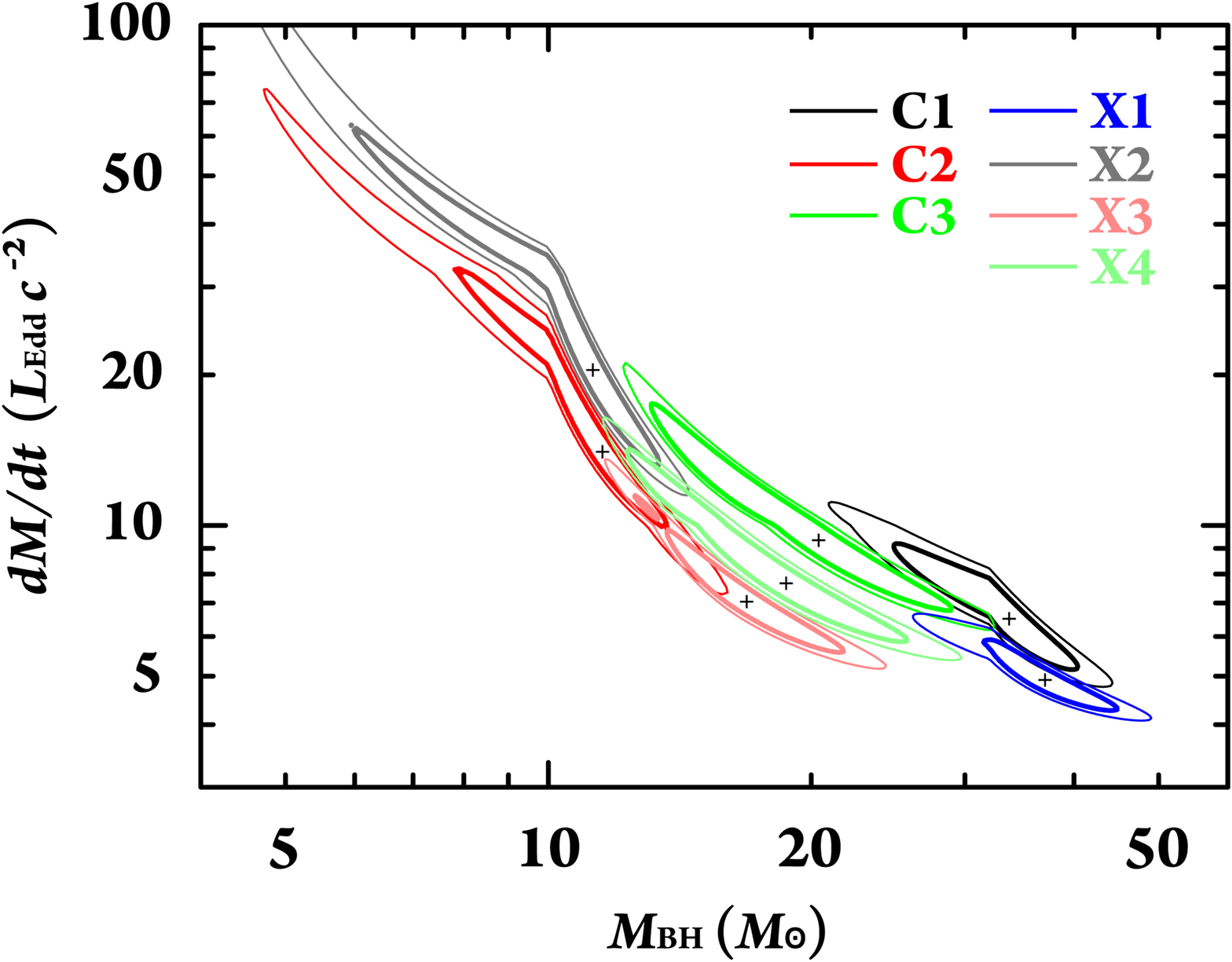}
\caption{Significance contours of the best-fit physical parameters ($dM/dt$ versus $M_{\rm{BH}}$) by the Kawaguchi model for M\,51 source-82;
The results in different observations are shown with different colors: 
C1 ({\it black}), C2 ({\it red}), C3 ({\it green}), X1 ({\it blue}), X2 ({\it gray}), X3 ({\it light red}), and X4 ({\it light green}).
The two confidence levels are shown for 68\% ({\it {thick solid}}) and 90\% ({\it {thin solid}}) ranges.
\label{fig8}}
\end{figure}

\subsection{Spectral Variations of Individual Sources}

Now, we look into details of individual sources.
The following result shows that the BH mass is in a range of 30--40~$M_{\odot}$ for sources which we can estimate a mass with a representative model.
This supports our assumption to derive the overall trend in $\S$~5.1.
All the samples can be considered to be in one of the three states defined in $\S$~5.1.
The spectra in the {\it{bright}} state is explained better with a curved model and those in the {\it{faint}} state with a PL model.
For the curved model, we see that the Kawaguchi model better explains the sources with sufficient constraints.

For each individual source, we discuss the states of all samples and the most likely representative model for each state.
The result of spectral fitting is employed, which is mainly used to distinguish PL-like or curved spectra.
If both the MCD and the Kawaguchi models give an acceptable fit for a curved spectrum,
we further use a plot of $L_{\rm{X,MCD}}$ against $kT_{\rm{in}}$ (Figure~\ref{fig6}) to distinguish the two.
The plot is commonly used for Galactic BHBs and ULXs.
The two parameters are correlated in a different manner in two different physical conditions: standard disk or slim disk.
If a system has a standard disk (the MCD model), the relation $L_{\rm{X,MCD}}$~$\propto$~${T_{\rm{in}}}^4$ is obtained,
which is interpreted that the innermost disk radius remains constant against varying accretion rates in the state \citep{sha73,ebi93,gie04}.
If a system has a slim disk (the Kawaguchi model), the relation $L_{\rm{X,MCD}}$~$\propto$~${T_{\rm{in}}}^{\beta}$ is obtained,
where $\beta$~$\lesssim$~4 \citep{wata00}.
The power $\beta$ decreases from 4 as an increasing mass accretion rate \citep{wata01}.
For samples with a successful fit by the MCD or Kawaguchi model or both,
we derived the best-fit $M_{\rm{BH}}$ value and examined if it is consistent among all samples belonging to a source.

\subsubsection{NGC\,4490/85 ULX-8}

We consider that ULX-8 exhibited an intra-state variation within the {\it{bright}} state.
This is because the luminosity variation (4.6--6.0$\times$10$^{39}$~ergs~s$^{-1}$) happens only above the border luminosity.

We speculate that this ULX is in the slim disk state based on the two lines of evidence.
The first is that the scatter in the $L_{\rm{X,MCD}}$--$T_{\rm{in}}$ plot (Figure~\ref{fig6}e)
follows $L_{\rm{X,MCD}}$~$\propto$~${T_{\rm{in}}}^{\beta}$, where $\beta$~$=$~1.0$^{+0.7}_{-0.4}$.
The value is consistent with the slim disk ($\beta$~$\lesssim$~4) and not with the standard disk ($\beta$~$\sim$~4).
We note that the value is not an apparent slope caused by an additional spectral component (e.g. an additional PL component generally found in the HSS),
because we obtained an even flatter slope ($\beta$~$=$~0.6$^{+0.7}_{-0.3}$) when the spectra were fitted by two component models (the PL plus the MCD).
Here, we fixed the photon index of the additional PL component to be 2.0 for the lack of statistics to constrain the value.

The second evidence is that a BH mass consistent among all the samples is found for the Kawaguchi model, but not for the MCD model.
Figure~\ref{fig7} shows the best-fit parameters using the MCD and Kawaguchi models.
In the MCD fitting (Figure~\ref{fig7}a),
a value ($R_{\rm{in}}$/km)$^2$/($D$/10~kpc)$^2$ $(\cos{i})^{-1}$~$\propto$~${M_{\rm{BH}}}^2$
consistent among the four samples was not found within 90\% statistical uncertainty, which is unphysical.
In other words, although spectra of ULXs in the {\it{bright}} state are commonly analyzed using the MCD model \citep[e.g.][]{kubo01},
it now turns out that the MCD model does not describe the long-term spectral variations of the {\it{bright}} state.
On the other hand, in the Kawaguchi model fitting (Figure~\ref{fig7}b),
the range ${M_{\rm{BH}}}$~$=$~35--38~$M_{\odot}$ was found to be consistent within 90\% among the four samples.
Thus, at least for this source, the Kawaguchi model is a more favorable description for its spectral variation.

Now we found a common BH mass range acceptable for the four samples,
we fitted the four spectra by tying the BH mass to obtain a more stringent constraint.
The best-fit value of ${M_{\rm{BH}}}$ is $37 \pm 2$~$M_{\odot}$ (Table~\ref{table6}).
Resultantly, the Eddington ratio ($L_{\rm{X}}$/$L_{\rm{Edd}}$) is 0.41--0.63,
which is similar to those of other ULX spectra considered to be in the slim disk state \citep[0.36 for NGC\,1313 X-2 and 0.52 for NGC\,4559 X-7;][]{kiki06}.

\begin{deluxetable}{cccc}
\tablecaption{Best-fit parameters of the Kawaguchi model for ULX-8.\label{table6}}
\tablewidth{0pt}
\tablecolumns{4}
\tablehead{
Data & $N_{\rm{H}}^{\prime}$ & $dM/dt$ & $L_{\rm{X,Kaw}}$ \\
label & $(10^{22}\;\rm{cm}^{-2})$ & $(L_{\rm{Edd}} \; c^{-2})$ & $(10^{38}\;\rm{ergs \; s^{-1}})$
}
\startdata
C1$^{\prime}$ & $ 0.68^{+0.10}_{-0.09} $ & $ 12^{+2}_{-1} $ & $ 30 \pm 3 $ \\
C2$^{\prime}$ & $ 0.55^{+0.07}_{-0.06} $ & $ 10^{+1}_{-1} $ & $ 22^{+\phantom{1}3}_{-\phantom{1}2} $ \\
C3$^{\prime}$ & $ 0.68^{+0.06}_{-0.06} $ & $ 15^{+2}_{-1} $ & $ 36 \pm 2 $ \\
X1$^{\prime}$ & $ 0.38^{+0.05}_{-0.04} $ & $ 11^{+1}_{-1} $ & $ 26^{+\phantom{1}3}_{-\phantom{1}2} $ \\
\tableline
\multicolumn{2}{l}{$M_{\rm{BH}}$\tablenotemark{a} $(M_{\odot})$} & & $ 37 \pm 2 $ \\
\multicolumn{2}{l}{Red$-\chi^2$ (d.o.f.)} & & 1.04(143)\phantom{1}
\enddata
\tablenotetext{a}{The BH mass is fixed among the four independent fittings.}
\end{deluxetable}

\subsubsection{M\,51 Source-82}

Source-82 shows variability among samples, which we consider to be an intra-state origin within the {\it{faint}} state.
This is because the luminosity variation (1.6--2.9$\times$10$^{39}$~ergs~s$^{-1}$) happens only below the border luminosity.

We consider that the source stayed in the state represented by the PL model in all samples based on the following three reasons:
(i) The majority of samples (six out of all seven) are successfully explained by a PL model.
(ii) None of the samples is explained by an MCD model.
(iii) Although the Kawaguchi model reproduces all the samples,
the model is unlikely because we do not find $M_{\rm{BH}}$ consistent among all the samples (Figure~\ref{fig8}).

In one sample (X1), source-82 shows a steep photon index of 2.6$^{+0.3}_{-0.2}$ (Table~\ref{table4}).
In contrast, in the other six samples, the derived index (${\it{\Gamma}}$~$\sim$~1.8--2.2; Table~\ref{table4}) is flatter than that in X1.

\subsubsection{M\,51 Source-69}

Source-69 clearly caused an inter-state transition.
This ULX exhibits the largest luminosity variation by at least a factor of 70,
and shows a transition among the {\it{bright}} state (X2, X3, and X4; 4.0--5.8$\times$10$^{39}$~ergs~s$^{-1}$), 
the {\it{faint}} state (C1 and C3; 2.6--2.9$\times$10$^{39}$~ergs~s$^{-1}$),
and the {\it{dim}} state (C2 and X1; $<$1.2$\times$10$^{39}$~ergs~s$^{-1}$).
 
We discuss the representative model for the three states of the ULX.
First, in the {\it{bright}} state, none of the PL, MCD, and Kawaguchi models explains all the three samples (Table~\ref{table4}).
In particular, two samples (X3 and X4) were not reproduced by any models.
We employed two component models (the PL plus the MCD or the Kawaguchi model), which did not improve the fit.
The representative model was not found for this source in this state.

Second, the PL is the representative model for the {\it{faint}} state, because it is the only model to reproduce spectra of all samples.
The derived index is quite flat of ${\it{\Gamma}}$~$<$~1.5.

Finally, for the {\it{dim}} state, any models can reproduce the spectra and the representative model was not constrained.
We note, however, that the spectra in this state is softer (${\it{\Gamma}}$~$\sim$~1.9--3.1) than in the other two states (${\it{\Gamma}}$~$<$~1.7).

\subsubsection{NGC\,4490/85 ULX-6}

We consider that ULX-6 caused an inter-state transition because the variation spans the three different luminosity ranges as M\,51 source-69:
the {\it{bright}} state (C3$^{\prime}$ and X1$^{\prime}$; 4.4--5.3$\times$10$^{39}$~ergs~s$^{-1}$),
the {\it{faint}} state (C1$^{\prime}$; $\sim$2.3$\times$10$^{39}$~ergs~s$^{-1}$),
and the {\it{dim}} state (C2$^{\prime}$; $\sim$7.2$\times$10$^{38}$~ergs~s$^{-1}$).

For the {\it{bright}} state, both the MCD and the Kawaguchi models can explain all the spectra,
and yielded a BH mass consistent among the two samples.
However, the flat slope ($\beta$~$=$~0.4$^{+1.7}_{-0.6}$) of the $L_{\rm{X,MCD}}$--$T_{\rm{in}}$ plot (Figure~\ref{fig6}d)
favors the slim disk interpretation ($\beta$~$\lesssim$~4).
Thus, the representative model of the ULX in this state is expected to be the Kawaguchi model.
We tried a simultaneous fitting by tying the $M_{\rm{BH}}$ value using the Kawaguchi model, and obtained the best-fit BH mass of $33^{+2}_{-5}$~$M_{\odot}$.
The Eddington ratio is 0.53--0.55.

For the {\it{faint}} and {\it{dim}} states, all the models can explain all the samples, thus the representative model was not constrained.
In the {\it{dim}} state, the PL fitting yielded the best-fit power of $\sim$~2.6,
which is similar to that found in the {\it{dim}} state of M\,51 source-69.

\subsubsection{NGC\,4490/85 the other ULXs}

The remaining three sources have no clear features, thus the representative model and the BH mass can hardly be constrained.
We briefly describe these sources.

Both ULX-2 and ULX-3 show intra-state variation as their luminosity
($L_{\rm{X,PL}}$~$\sim$~3.3--5.2 and 3.5--5.8$\times$10$^{39}$~ergs~s$^{-1}$, respectively) stays in the {\it{bright}} state.
For ULX-2, no model explains all the samples, thus the representative model was not found for this source.
For ULX-3, any models can reproduce all the spectra.
The $L_{\rm{X,MCD}}$--$T_{\rm{in}}$ plot does not separate the slim disk or the standard disk state,
because the data cannot be fitted with a linear relation.
A BH mass consistent among all the samples is found both for the MCD and the Kawaguchi models as $8 \pm 1$~$M_{\odot}$ and $28^{+6}_{-7}$~$M_{\odot}$, respectively.

ULX-4, on the other hand, is considered to have exhibited an inter-state transition.
The luminosities of the three samples are in the {\it{bright}} state (C2$^{\prime}$, C3$^{\prime}$, and X1$^{\prime}$; 4.2--5.1$\times$10$^{39}$~ergs~s$^{-1}$),
while that of the other is in the {\it{faint}} state (C1$^{\prime}$; $\sim$2.5$\times$10$^{39}$~ergs~s$^{-1}$).
In both states, any models explain most of the samples,
and the representative model cannot be constrained for the same reasons as ULX-3.
If we estimate the BH mass using a simultaneous fitting of the two samples in the {\it{bright}} state,
the BH mass is obtained as $9 \pm 1$~$M_{\odot}$ (MCD) and $26^{+\phantom{1}6}_{-10}$~$M_{\odot}$ (Kawaguchi).

\subsection{Relations with Galactic BHBs}

Many previous studies \citep[e.g.][]{kubo01,dew05,win06,kiki06,glad09,god09} suggested that the states of ULXs are related to those of Galactic BHBs.
In the present study, we uniformly analyzed 34 spectra of seven ULXs in two interacting galaxies.
As shown in $\S$~5.2, we found some patterns of luminosity and spectral variations.
We proceed to compare each state of the ULXs with those of Galactic BHBs.

For the {\it{bright}} state, especially for NGC\,4490/85 ULX-8 and 6,
we speculate that this state corresponds to the ASS based on the following two reasons:
(i) The two sources with the slim disk are the most likely interpretation ($\S$~5.2.1 and 5.2.4).
(ii) The Eddington ratios (0.41--0.63 and 0.53--0.55) are similar to those of two Galactic BHBs considered to be in the ASS:
XTE\,J1550--564 \citep[$\sim$0.4;][]{kubo04} and 4U\,1630--47 \citep[0.25--0.35;][]{abe05}.
However, the {\it{bright}} state of M\,51 source-69 and NGC\,4490/85 ULX-2 do not show such features of the ASS.
This fact suggests that the boundary luminosity to separate the bright and the faint states may be correct overall, but varies source to source.

For the {\it{faint}} state, the representative model is PL for sources which we can determine a model (M\,51 source-82 and 69).
The VHS and the LHS in Galactic BHBs have a PL shaped spectrum with the photon index ${\it{\Gamma}}$ of $\sim$2.5 and 1.5--1.9, respectively.
However, our samples in the {\it{faint}} state show a wide PL index range of ${\it{\Gamma}}$~$\sim$~1.1--2.6.
Because a state with such a steep index of 1.1 has not been known so far,
the {\it{faint}} state may be different from the well-known state in Galactic BHBs.

Finally, we briefly mention the {\it{dim}} state.
We consider the {\it{dim}} state seen in some sources is distinctive from the {\it{faint}} states.
This is because this state has lower luminosity and steeper spectral index, when fitted with PL, than the faint state.
The photon statistics is too low to constrain the representative model, and their counterpart state in the Galactic BHBs is unknown.

\section{Summary and Conclusions}

We have analyzed X-ray spectra (34~samples) of seven bright ULXs in the interacting galaxy systems M\,51 and NGC\,4490/85
using the archived data from multiple {\it{Chandra}} and {\it{XMM-Newton}} observations.
We constructed a histogram of luminosities, in which we found a hint of three distinctive states:
(i) The {\it{bright}} state with the brighter luminosity range (3--6$\times$10$^{39}$~ergs~s$^{-1}$) for 20~samples,
(ii) the {\it{faint}} state with the fainter luminosity range (1.5--3$\times$10$^{39}$~ergs~s$^{-1}$) for 11~samples,
and (iii) the {\it{dim}} state with luminosities below the fainter range ($<$1.2$\times$10$^{39}$~ergs~s$^{-1}$) for three samples.
On the whole, the spectra in the {\it{bright}} state are represented by the MCD model (standard disk) or the Kawaguchi model (slim disk),
while those in the {\it{faint}} state are represented by the PL model.
We classified the flux changes of all sources into the intra-state variation and the inter-state transition.
Four ULXs (M\,51 source-82, NGC\,4490/85 ULX-2, 3, and 8) exhibited the former,
while the remaining three (M\,51 source-69, NGC\,4490/85 ULX-4, and 6) exhibited the latter.

All the spectra of two ULXs (NGC\,4490/85 ULX-8 and 6) in the {\it{bright}} state
can be reproduced by a slim disk model by \citet{kawa03} with a constant BH mass of $37 \pm 2$ and $33^{+2}_{-5}$~$M_{\odot}$, respectively.
In particular, for ULX-8, a common BH mass was not found with the MCD model fitting.
This suggests that the slim disk model is a more appropriate explanation at least in some cases of the ``{\it{bright}}'' state,
in which ULXs harbor BHs of a mass of $<$40~$M_{\odot}$.

From the results of spectral fitting, we compare each state of the ULXs with those of Galactic BHBs.
We propose that the {\it{bright}} state of two ULXs (NGC\,4490/85 ULX-8 and 6) corresponds to the ASS of Galactic BHBs.
For the {\it{faint}} state, the representative model is PL.
However, it is different in the range of photon index from the VHS of Galactic BHBs, which also shows a PL spectrum.
We thus consider that the {\it{faint}} state is unique to ULXs.
The nature of the {\it{dim}} state is unconstrained, but has the most steep spectrum.

\acknowledgments

We thank the anonymous reviewer for critical comments for the paper.
This research has made use of public data and software obtained from
the High Energy Astrophysics Science Archive Research Center (HEASARC), provided by NASA's Goddard Space Flight Center.
Funding for the SDSS and SDSS-II has been provided by the Alfred P.~Sloan Foundation,
the Participating Institutions, the National Science Foundation, the U.S. Department of Energy,
the National Aeronautics and Space Administration, the Ministry of Education, Culture, Sports, Science, and Technology of Japan,
the Max Planck Society, and the Higher Education Funding Council for England.
The SDSS Web Site is http://www.sdss.org/.
SDSS images used in this paper was produced with Montage,
an image mosaic service supported by the NASA Earth Sciences Technology Office Computing Technologies program.
Tessei Yoshida is supported by the Japan Society for the Promotion of Science Research Fellowship for Young Scientists.

{\it Facilities:} \facility{CXO (ACIS)}, \facility{XMM (EPIC)}.


\begin{thebibliography}{}
\bibitem[Abe et al.(2005)]{abe05} Abe, Y., Fukazawa, Y., Kubota, A., Kasama, D., \& Makishima, K.\ 2005, \pasj, 57, 629
\bibitem[Abramowicz et al.(1988)]{abr88} Abramowicz, M.~A., Czerny, B., Lasota, J.~P., \& Szuszkiewicz, E.\ 1988, \apj, 332, 646
\bibitem[Brassington et al.(2007)]{bra07} Brassington, N.~J., Ponman, T.~J., \& Read, A.~M.\ 2007, \mnras, 377, 1439 
\bibitem[Broos et al.(2002)]{bro02} Broos, P.~S., Townsley, L.~K., Getman, K., \& Bauer, F.~E.\ 2002, ACIS Extract, An ACIS Point Source Extraction Package (University Park: Pennsylvania State Univ.)
\bibitem[Clemens et al.(1998)]{clem98} Clemens, M.~S., Alexander, P., \& Green, D.~A.\ 1998, \mnras, 297, 1015 
\bibitem[Clemens et al.(1999)]{clem99} Clemens, M.~S., Alexander, P., \& Green, D.~A.\ 1999, \mnras, 307, 481
\bibitem[Clemens, \& Alexander(2002)]{clem02} Clemens, M.~S., \& Alexander, P.\ 2002, \mnras, 333, 39
\bibitem[Davis et al.(2006)]{dav06} Davis, S.~W., Done, C., \& Blaes, O.~M.\ 2006, \apj, 647, 525
\bibitem[Dewangan et al.(2005)]{dew05} Dewangan, G.~C., Griffiths, R.~E., Choudhury, M., Miyaji, T., \& Schurch, N.~J.\ 2005, \apj, 635, 198 
\bibitem[Dickey, \& Lockman(1990)]{dic90} Dickey, J.~M., \& Lockman, F.~J.\ 1990, \araa, 28, 215
\bibitem[de Vaucouleurs et al.(1976)]{deVau76} de Vaucouleurs, G., de Vaucouleurs, A., \& Corwin, J.~R.\ 1976, Second Reference Catalogue of Bright Galaxies (Austin: Univ. Texas Press)
\bibitem[de Vaucouleurs et al.(1991)]{deVau91} de Vaucouleurs, G., de Vaucouleurs, A., Corwin, H.~G., Jr., Buta, R.~J., Paturel, G., \& Fouque, P. \ 1991, Third Reference Catalogue of Bright Galaxies (New York: Springer)
\bibitem[Ebisawa et al.(1993)]{ebi93} Ebisawa, K., Makino, F., Mitsuda, K., Belloni, T., Cowley, A.~P., Schmidtke, P.~C., \& Treves, A.\ 1993, \apj, 403, 684
\bibitem[Ebisawa et al.(1994)]{ebi94} Ebisawa, K., et al.\ 1994, \pasj, 46, 375
\bibitem[Ebisawa et al.(2003)]{ebi03} Ebisawa, K., {\.Z}ycki, P., Kubota, A., Mizuno, T., \& Watarai, K.\ 2003, \apj, 597, 780
\bibitem[Ehle et al.(1995)]{ehle95} Ehle, M., Pietsch, W., \& Beck, R.\ 1995, \aap, 295, 289
\bibitem[Esin et al.(1997)]{esin97} Esin, A.~A., McClintock, J.~E., \& Narayan, R.\ 1997, \apj, 489, 865
\bibitem[Fabbiano(2006)]{fab06} Fabbiano, G.\ 2006, \araa, 44, 323
\bibitem[Feldmeier et al.(1997)]{fel97} Feldmeier, J.~J., Ciardullo, R., \& Jacoby, G.~H.\ 1997, \apj, 479, 231 
\bibitem[Feng \& Kaaret(2010)]{fen10} Feng, H., \& Kaaret, P.\ 2010, \apjl, 712, L169
\bibitem[Foschini et al.(2006)]{fos06} Foschini, L., et al.\ 2006, Advances in Space Research, 38, 1378
\bibitem[Fridriksson et al.(2008)]{fri08} Fridriksson, J.~K., Homan, J., Lewin, W.~H.~G., Kong, A.~K.~H., \& Pooley, D.\ 2008, \apjs, 177, 465
\bibitem[Fukazawa et al.(2001)]{fuk01} Fukazawa, Y., Iyomoto, N., Kubota, A., Matsumoto, Y., \& Makishima, K.\ 2001, \aap, 374, 73 
\bibitem[Garmire et al.(2003)]{gar03} Garmire, G.~P., Bautz, M.~W., Ford, P.~G., Nousek, J.~A., \& Ricker, G.~R., Jr.\ 2003, \procspie, 4851, 28
\bibitem[Gierli{\'n}ski, \& Done(2004)]{gie04} Gierli{\'n}ski, M., \& Done, C.\ 2004, \mnras, 347, 885
\bibitem[Gladstone, \& Roberts(2009)]{glad09} Gladstone, J.~C., \& Roberts, T.~P.\ 2009, \mnras, 397, 124
\bibitem[Goad et al.(2006)]{goa06} Goad, M.~R., Roberts, T.~P., Reeves, J.~N., \& Uttley, P.\ 2006, \mnras, 365, 191
\bibitem[Godet et al.(2009)]{god09} Godet, O., Barret, D., Webb, N.~A., Farrell, S.~A., \& Gehrels, N.\ 2009, \apjl, 705, L109
\bibitem[Jansen et al.(2001)]{jan01} Jansen, F., et al.\ 2001, \aap, 365, L1
\bibitem[Kaaret et al.(2001)]{kaa01} Kaaret, P., Prestwich, A.~H., Zezas, A., Murray, S.~S., Kim, D.-W., Kilgard, R.~E., Schlegel, E.~M., \& Ward, M.~J.\ 2001, \mnras, 321, L29
\bibitem[Kaaret et al.(2004)]{kaa04} Kaaret, P., Ward, M.~J., \& Zezas, A.\ 2004, \mnras, 351, L83
\bibitem[Kawaguchi(2003)]{kawa03} Kawaguchi, T.\ 2003, \apj, 593, 69
\bibitem[King et al.(2001)]{king01} King, A.~R., Davies, M.~B., Ward, M.~J., Fabbiano, G., \& Elvis, M.\ 2001, \apjl, 552, L109
\bibitem[Kubota et al.(1998)]{kubo98} Kubota, A., Tanaka, Y., Makishima, K., Ueda, Y., Dotani, T., Inoue, H., \& Yamaoka, K.\ 1998, \pasj, 50, 667 
\bibitem[Kubota et al.(2001)]{kubo01} Kubota, A., Mizuno, T., Makishima, K., Fukazawa, Y., Kotoku, J., Ohnishi, T., \& Tashiro, M.\ 2001, \apjl, 547, L119
\bibitem[Kubota, \& Makishima(2004)]{kubo04} Kubota, A., \& Makishima, K.\ 2004, \apj, 601, 428
\bibitem[Kuno, \& Nakai(1997)]{kuno97} Kuno, N., \& Nakai, N.\ 1997, \pasj, 49, 279
\bibitem[Liu et al.(2002)]{liu02} Liu, J.-F., Bregman, J.~N., Irwin, J., \& Seitzer, P.\ 2002, \apjl, 581, L93
\bibitem[Liu, \& Bregman(2005)]{liu05} Liu, J.-F., \& Bregman, J.~N.\ 2005, \apjs, 157, 59
\bibitem[Maccarone(2003)]{mac03} Maccarone, T.~J.\ 2003, \aap, 409, 697
\bibitem[Makishima et al.(2000)]{max00} Makishima, K., et al.\ 2000, \apj, 535, 632
\bibitem[Marston et al.(1995)]{mar95} Marston, A.~P., Elmegreen, D., Elmegreen, B., Forman, W., Jones, C., \& Flanagan, K.\ 1995, \apj, 438, 663
\bibitem[Matsumoto et al.(2001)]{mat01} Matsumoto, H., Tsuru, T.~G., Koyama, K., Awaki, H., Canizares, C.~R., Kawai, N., Matsushita, S., \& Kawabe, R.\ 2001, \apjl, 547, L25 
\bibitem[McClintock et al.(2007)]{mcc07} McClintock, J.~E., Narayan, R., \& Shafee, R.\ 2007, Black Holes (Cambridge: Cambridge Univ. Press), arXiv:0707.4492 
\bibitem[Miller, \& Colbert(2004)]{mil04} Miller, M.~C., \& Colbert, E.~J.~M.\ 2004, International Journal of Modern Physics D, 13, 1
\bibitem[Mitsuda et al.(1984)]{mit84} Mitsuda, K., et al.\ 1984, \pasj, 36, 741 
\bibitem[Miyawaki et al.(2009)]{miy09} Miyawaki, R., Makishima, K., Yamada, S., Gandhi, P., Mizuno, T., Kubota, A., Tsuru, T.~G., \& Matsumoto, H.\ 2009, \pasj, 61, 263 
\bibitem[Mizuno et al.(2007)]{miz07} Mizuno, T., et al.\ 2007, \pasj, 59, 257 
\bibitem[Morrison, \& McCammon(1983)]{mor83} Morrison, R., \& McCammon, D.\ 1983, \apj, 270, 119
\bibitem[Ohsuga et al.(2005)]{ohs05} Ohsuga, K., Mori, M., Nakamoto, T., \& Mineshige, S.\ 2005, \apj, 628, 368
\bibitem[Okajima et al.(2006)]{oka06} Okajima, T., Ebisawa, K., \& Kawaguchi, T.\ 2006, \apjl, 652, L105
\bibitem[Palumbo et al.(1985)]{pal85} Palumbo, G.~G.~C., Fabbiano, G., Trinchieri, G., \& Fransson, C.\ 1985, \apj, 298, 259
\bibitem[Pringle(1981)]{pri81} Pringle, J.~E.\ 1981, \araa, 19, 137
\bibitem[Ptak et al.(1999)]{ptak99a} Ptak, A., Serlemitsos, P., Yaqoob, T., \& Mushotzky, R.\ 1999, \apjs, 120, 179
\bibitem[Ptak, \& Griffiths(1999)]{ptak99b} Ptak, A., \& Griffiths, R.\ 1999, \apjl, 517, L85
\bibitem[Read et al.(1997)]{read97} Read, A.~M., Ponman, T.~J., \& Strickland, D.~K.\ 1997, \mnras, 286, 626 
\bibitem[Roberts, \& Warwick(2000)]{rob00} Roberts, T.~P., \& Warwick, R.~S.\ 2000, \mnras, 315, 98
\bibitem[Roberts et al.(2002)]{rob02} Roberts, T.~P., Warwick, R.~S., Ward, M.~J., \& Murray, S.~S.\ 2002, \mnras, 337, 677
\bibitem[Roberts et al.(2006)]{rob06} Roberts, T.~P., Kilgard, R.~E., Warwick, R.~S., Goad, M.~R., \& Ward, M.~J.\ 2006, \mnras, 371, 1877
\bibitem[Schweizer(1977)]{sch77} Schweizer, F.\ 1977, \apj, 211, 324
\bibitem[Scoville et al.(2001)]{sco01} Scoville, N.~Z., Polletta, M., Ewald, S., Stolovy, S.~R., Thompson, R., \& Rieke, M.\ 2001, \aj, 122, 3017
\bibitem[Shakura, \& Sunyaev(1973)]{sha73} Shakura, N.~I., \& Sunyaev, R.~A.\ 1973, \aap, 24, 337
\bibitem[Shimura, \& Takahara(1995)]{sim95} Shimura, T., \& Takahara, F.\ 1995, \apj, 445, 780
\bibitem[Str{\"u}der et al.(2001)]{stru01} Str{\"u}der, L., et al.\ 2001, \aap, 365, L18
\bibitem[Terashima et al.(1998)]{tera98} Terashima, Y., Ptak, A., Fujimoto, R., Itoh, M., Kunieda, H., Makishima, K., \& Serlemitsos, P.~J.\ 1998, \apj, 496, 210
\bibitem[Terashima, \& Wilson(2004)]{tera04} Terashima, Y., \& Wilson, A.~S.\ 2004, \apj, 601, 735
\bibitem[Terashima et al.(2006)]{tera06} Terashima, Y., Inoue, H., \& Wilson, A.~S.\ 2006, \apj, 645, 264
\bibitem[Tsunoda et al.(2006)]{tsuno06} Tsunoda, N., Kubota, A., Namiki, M., Sugiho, M., Kawabata, K., \& Makishima, K.\ 2006, \pasj, 58, 1081
\bibitem[Turner et al.(2001)]{tur01} Turner, M.~J.~L., et al.\ 2001, \aap, 365, L27
\bibitem[van der Hulst et al.(1988)]{van88} van der Hulst, J.~M., Kennicutt, R.~C., Crane, P.~C., \& Rots, A.~H.\ 1988, \aap, 195, 38
\bibitem[V{\'a}zquez et al.(2007)]{vaz07} V{\'a}zquez, G.~A., Hornschemeier, A.~E., Colbert, E., Roberts, T.~P., Ward, M.~J., \& Malhotra, S.\ 2007, \apjl, 658, L21
\bibitem[Viallefond et al.(1980)]{via80} Viallefond, F., Allen, R.~J., \& de Boer, J.~A.\ 1980, \aap, 82, 207
\bibitem[Vierdayanti et al.(2006)]{kiki06} Vierdayanti, K., Mineshige, S., Ebisawa, K., \& Kawaguchi, T.\ 2006, \pasj, 58, 915 
\bibitem[Watarai et al.(2000)]{wata00} Watarai, K., Fukue, J., Takeuchi, M., \& Mineshige, S.\ 2000, \pasj, 52, 133
\bibitem[Watarai et al.(2001)]{wata01} Watarai, K., Mizuno, T., \& Mineshige, S.\ 2001, \apjl, 549, L77
\bibitem[Weisskopf et al.(2002)]{wei02} Weisskopf, M.~C., Brinkman, B., Canizares, C., Garmire, G., Murray, S., \& Van Speybroeck, L.~P.\ 2002, \pasp, 114, 1
\bibitem[Winter et al.(2006)]{win06} Winter, L.~M., Mushotzky, R.~F., \& Reynolds, C.~S.\ 2006, \apj, 649, 730
\end{thebibliography}
\end{document}